\documentclass[5p]{elsarticle} 

\usepackage{lineno}
\modulolinenumbers[5]

\journal{NIM A}
\usepackage{graphicx,xcolor,textpos,tabularx}
\usepackage{epstopdf}
\usepackage{amsmath}
\usepackage{xcolor}
\usepackage{import}
\usepackage{graphicx} 
\usepackage{dcolumn}  
\usepackage{bm}       
\usepackage{amssymb}
\usepackage[colorlinks=true, allcolors=blue]{hyperref}

\usepackage{cuted}

\biboptions{longnamesfirst,sort&compress}

\begin{document}
\graphicspath{{./Figures/}}
\begin{frontmatter}

\title{A high-voltage MR-ToF mass spectrometer and separator for the study of exotic isotopes at FRIB
}


\author[FRIB]{F.M.~Maier}
\corref{mycorrespondingauthor}
\cortext[mycorrespondingauthor]{Corresponding author}
\ead{maierf@frib.msu.edu}
\author[FRIB,MSU]{C.M.~Ireland}
\author[FRIB,MSU]{G.~Bollen}
\author[FRIB,MSU]{E. Dhayal}
\author[Ed]{T. Fowler-Davis}
\author[Berkeleyadress]{E.~Leistenschneider}
\author[Ed]{M.P.~Reiter}
\author[FRIB,MSU]{R.~Ringle}
\author[FRIB]{S.~Schwarz}
\author[FRIB,MSU2]{A.~Sjaarda}

\address[FRIB]{Facility for Rare Isotope Beams, East Lansing, Michigan 48824, USA}
\address[MSU]{Department of Physics and Astronomy, Michigan State University, East Lansing, Michigan 48824, USA}
\address[Ed]{School of Physics and Astronomy, University of Edinburgh, Edinburgh EH9 3FD, UK}
\address[Berkeleyadress]{Nuclear Science Division, Lawrence Berkeley National Laboratory, Berkeley, California 94720, USA.}
\address[MSU2]{College of Engineering, Michigan State University, East Lansing, Michigan 48824, USA}


\begin{abstract}
The Facility for Rare Isotope Beams (FRIB) delivers a wide variety of rare isotopes as fast, stopped, or reaccelerated beams to enable forefront research in nuclear structure, astrophysics, and fundamental interactions. To expand the scientific potential of FRIB’s stopped and reaccelerated beam programs, we are designing a Multi-Reflection Time-of-Flight mass spectrometer and separator (MR-ToF MS).  It will enable high-precision mass measurements of short-lived isotopes, improve beam diagnostics, and deliver isobarically and isomerically purified beams to downstream experimental stations. It is designed to store ions at a kinetic energy of 30 keV, significantly enhancing ion throughput while maintaining high mass resolving power. We present the scientific motivation, technical design, and simulations demonstrating the expected performance of the system, which has the potential to significantly enhance FRIB's mass measurement, diagnostic, and mass separation capabilities. 
\end{abstract}

\end{frontmatter}
\section{\label{sec:Introduction}Introduction}
The Facility for Rare Isotope Beams (FRIB) is a next-generation rare-isotope facility that produces exotic short-lived radionuclides far from stability. It began delivering beams to experiments in 2022. Many of the nuclides have never been studied before, and their unique behaviors challenge existing theoretical models and open the door to new physical insights. Measuring their properties deepens our understanding of nuclear structure, astrophysics, and fundamental interactions.  Experiments at FRIB can be performed with fast ($\approx$ 200 MeV/u), stopped (0 eV up to 60~keV\footnote{Note that all beam energies stated throughout this manuscript are given for singly charged ions.}) and reaccelerated (300 keV/u to 12 MeV/u) rare-isotope beams~\cite{VILLARI2023350, SUMITHRARACHCHI2020305, LUND2020378}. 

To further expand the scientific opportunities with FRIB's stopped and reaccelerated beams, we are designing a Multi-Reflection Time-of-Flight mass spectrometer and separator (MR-ToF MS). It is expected to enhance the reach of FRIB's high-precision mass measurement program, deliver isobarically and isomerically purified ion beams to experimental stations, and improve beam diagnostics and identification. Over the past 15 years, MR-ToF MSs have gained remarkable success in their use at rare-isotope facilities around the world~\cite{PLASS2013, Schury2009, Wienholtz2013, REITER2021165823, CHAUVEAU2016211,LIU2021164679, ROSENBUSCH2022167824, virtanen2025highresolutionmultireflectiontimeofflightmass}. 
They typically consist of a Paul-trap cooler buncher preparing the ions for optimal injection into the MR-ToF device, which is made out of two opposing electrostatic mirrors that surround a central drift tube. As ions are being reflected back and forth isochronously between the mirrors, they separate in time-of-flight according to their mass-to-charge ratio ($m/q$).  After being ejected from the MR-ToF device, the ions impinge on a retractable ion detector that registers their arrival time. As ions with different $m/q$ travel at different velocities, they arrive at distinct times, forming well-separated peaks in the time-of-flight spectrum. Each peak corresponds to a specific $m/q$ ratio, enabling identification and precise mass measurements of the isotope of interest relative to a reference (atomic or molecular) ion with a well-known mass. 
In addition to mass measurements and beam diagnostics, MR-ToF MSs are also used to provide purified ion beams for subsequent experiments~\cite{ Wienholtz2013, DICKEL2015172,REITER2021165823,CHAUVEAU2016211}. When operated in mass separation mode, the ion detector is retracted, and the ions pass through a beam gate~\cite{PhysRev.49.388, PLA20084560, WOLF201282} installed downstream of the MR-ToF MS that is transparent only during the time window when the ions of interest are passing. In this way, a purified ion beam can be provided to downstream experimental stations.

FRIB's MR-ToF device is intended to store ions at an unprecedented beam energy of 30~keV. Simulations~\cite{MAIER2023168545} have shown that the ion throughput of MR-ToF MS can be significantly enhanced by increasing the kinetic energy of the stored ions, leading to a higher beam intensity of the purified ion beam. Furthermore, the increased beam energy can also result in a larger energy spread tolerance, enabling higher mass resolving and separation powers in shorter storage times~\cite{MAIER2023168545,YAVOR20181}. The mass resolving power refers to the ability to distinguish between two ion species with similar masses, while the mass separation power describes the ability to fully isolate and transmit only the desired ion species. Sec.~\ref{sec:Motivation} presents how this new MR-ToF MS can expand FRIB's ability to explore rare isotopes and accelerate scientific discoveries; Sec.~\ref{sec:Setup} provides a detailed overview of the proposed setup, and Sec.~\ref{sec:Performance} discusses its expected performance.

\section{\label{sec:Motivation}Purpose of FRIB's next generation MR-ToF MS}
\subsection{\label{sec: Expanding Mass Measurements} Expanding the reach of FRIB's high-precision mass measurement program}
The mass $m$ of a nucleus is one of its most fundamental observables. It provides a unique fingerprint of a nucleus, enabling unambiguous identification of a species when determined with sufficient precision. Beyond identification, it serves as a sensitive indicator of the underlying nuclear shell configuration and offers a benchmark for nuclear models, supporting a better prediction of properties near driplines, where experimental data remain scarce~\cite{YAMAGUCHI2021103882, Blaum2024}. Furthermore, mass measurements of radioactive isotopes far from stability have greatly enhanced our understanding of astrophysical processes in the universe and will continue to do so~\cite{Clark2023, mumpower2016impact}. 
At FRIB, high-precision mass measurements are currently performed using the 9.4~T Penning trap at the Low-Energy-Beam and Ion-Trap (LEBIT) facility~\cite{RINGLE201387} or with fast beams and a magnetic spectrometer using the time-of-flight-magnetic-rigidity (ToF-B$\rho$) technique~\cite{MATOS2012171,MEISEL2013145}. While Penning traps enable the highest resolving power $>10^7$ and highest precision with uncertainties in the mass excess $<$ 1~keV, their application is typically limited to isotopes with half-lives $\gtrapprox 50$~ms. Additionally, overwhelming isobaric contamination may prevent a Penning-trap mass measurement. The ToF-B$\rho$ technique enables mass measurements of isotopes with half-lives down to hundreds of nanoseconds, but typically only achieves a resolving power of about $10^4$ and a precision above 100~keV. This precision is insufficient for many scientific inquiries, and can be subject to unresolved isomeric states.

To complement existing capabilities and overcome these limitations, many rare-isotope facilities use MR-ToF MSs~\cite{PLASS2013, Schury2009, Wienholtz2013, REITER2021165823, CHAUVEAU2016211,LIU2021164679, ROSENBUSCH2022167824, cannarozzo2025isomericyieldratiosmass}. They allow the resolution of small mass differences $\Delta m$, achieving a mass resolving power $m/\Delta m > 10^5$ within a storage time of a few (tens of) milliseconds. This enables high-precision mass measurements with uncertainties in the mass excess down to around 10~keV also for nuclei with half-lives as low as a few ms.  While not reaching the mass resolving power and precision that Penning traps can provide for longer-lived nuclei, MR-ToF MS have shown to be at least an order of magnitude more sensitive~\cite{Dickel2024, PhysRevLett.124.092502, Mougeot2021, PhysRevLett.120.062503} than the Time-of-Flight Ion-Cyclotron-Resonance (ToF-ICR) technique~\cite{ToF1,BECKER199053,KONIG199595} and also more sensitive than the Phase-Imaging Ion-Cyclotron-Resonance (PI-ICR) technique~\cite{PhysRevLett.110.082501, Eliseev2014} that are in standard use for Penning-trap mass measurements with radioactive ions. Successful MR-ToF mass measurements with an uncertainty of $\approx 20$~keV in the mass excess have been performed with as few as 10 ion counts over the entire measurement time~\cite{PhysRevC.103.034319}.

\begin{figure}[t]
\centering
\includegraphics[width=1\columnwidth]{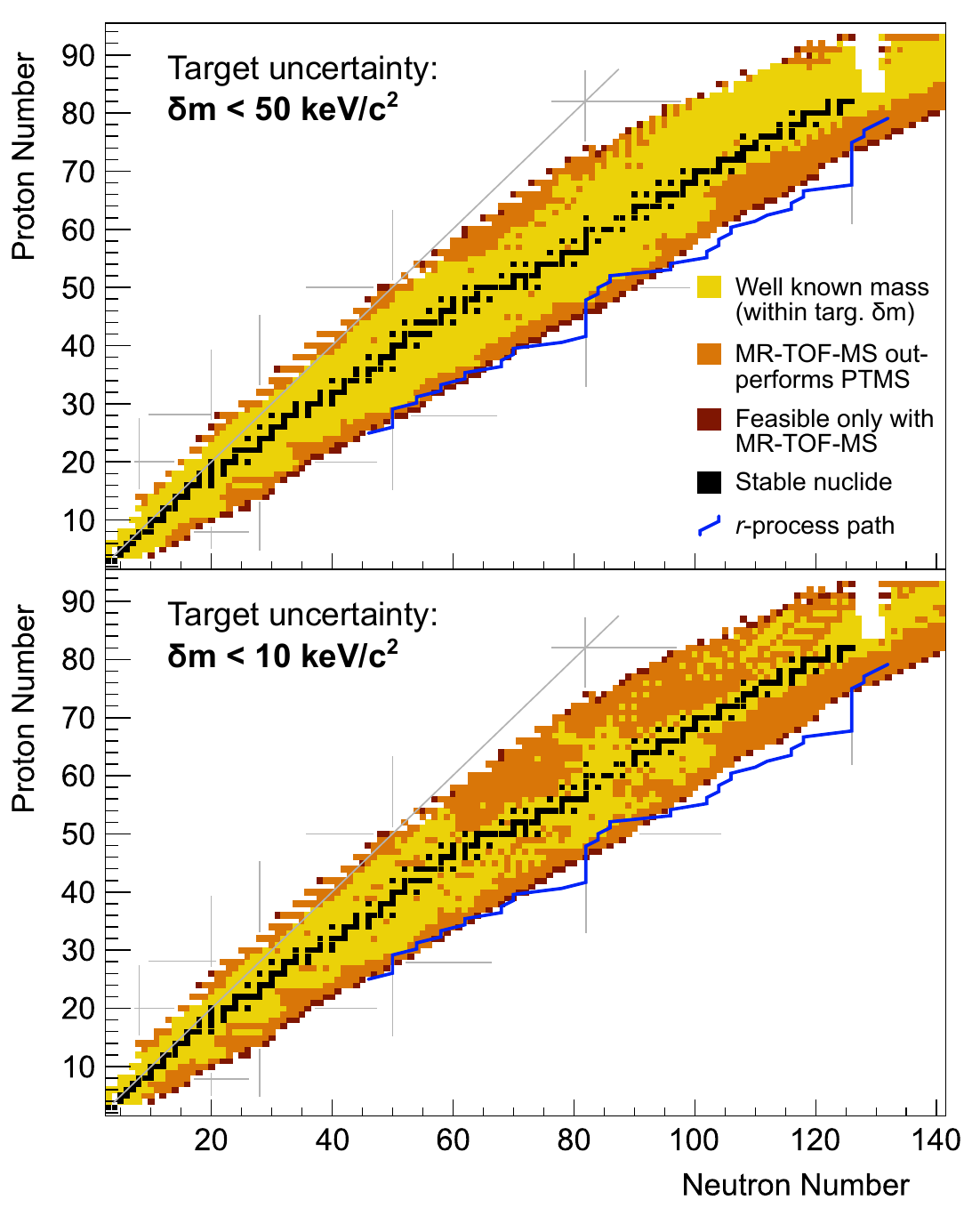}
\caption{Nuclei accessible for high-precision mass measurements within less than 60 hours of beam-on-target time at FRIB based on recent yield estimates for a 400~kW primary beam~\cite{Bradcom}. The top panel displays nuclei for which the mass uncertainty is below 50 keV, while the bottom panel highlights those with uncertainties below 10 keV. Stable nuclei are coloured in black, nuclei with a well-known mass in AME2020~\cite{AME2020} are depicted in yellow, nuclei for which a high-precision mass measurement is expected to be feasible at FRIB with either the LEBIT Penning trap or an MR-ToF MS, but ultimately with less beam-on-target time with the MR-TOF-MS, are shown in orange. Nuclei for which a high-precision mass measurement is foreseen to only be possible with an MR-ToF MS are shown in brown. The gray lines indicate self-conjugate nuclei with $N=Z$ and nuclei with a magic proton $Z$ or neutron number $N$, respectively. The weak r-process path shown in blue is taken from~\cite{annurev:/content/journals/10.1146/annurev.astro.46.060407.145207}.}
\label{fig: nuclear chart}
\end{figure}

 Fig.~\ref{fig: nuclear chart} illustrates the nuclei expected to be produced by FRIB over the coming decades and which masses could be measured by LEBIT's Penning trap and/or FRIB's MR-ToF MS within less than 60 hours of beam-on-target time assuming a 400~kW primary beam. Nuclei for which the target uncertainty of 50 keV (top panel) or 10 keV (bottom panel) has already been reported in the atomic mass evaluation AME2020~\cite{AME2020} are shown in yellow. Nuclei that are foreseen to be accessible at FRIB either with the Penning trap or the MR-ToF MS - though with less beam-on-target time with the MR-ToF MS - are shown in orange. Nuclei for which a mass measurement with the targeted precision will only be possible with the MR-ToF MS are marked in brown. The number of detected ions depends on the targeted precision of either 50 or 10 keV, but a minimum of 20 detected counts during the entire beam-on-target time is enforced. The measured total efficiency (including injection into the Paul-trap cooler-buncher and the intrinsic detection efficiency) for PI-ICR measurements at LEBIT is typically around 15\% without radioactive decay taken into account~\cite{PhysRevLett.132.152501}, while for the proposed MR-ToF MS we assume 50\%. If we assumed 15\% as the total efficiency for the MR-ToF MS as well, then the number of nuclei uniquely accessible by the MR-TOF MS would be reduced by ca. 25\%.  The cooling time in the Paul-trap cooler buncher is 5 ms as required for an optimal preparation of the ion bunch. The storage time in the Penning trap or MR-ToF MS depends on the targeted precision, but is at least 3~ms, enabling to reach a mass resolving power above $3\cdot10^5$ (see Fig.~\ref{fig: R sim} blue dashed curve for the proposed MR-ToF MS). According to Fig.~\ref{fig: nuclear chart}, we will be able to measure the mass of 890 nuclei with an uncertainty better than 50 keV at FRIB, which had not been measured with this uncertainty in AME2020. We will furthermore be able to improve the uncertainty of 1317 nuclei that are not known with an uncertainty better than 10 keV in AME2020. Out of these nuclei, 131 (for $<50$~keV uncertainty) or 119 (for $<10$~keV uncertainty) will be accessible only with an MR-ToF MS. However, also for the other nuclei less beam-on-target time is required for the MR-ToF MS to achieve the targeted uncertainty of either 50 or 10~keV on the mass excess.  

Furthermore, if higher measurement precision is required, or if unresolved isobars or isomers remain, the MR-ToF MS can, for nuclei with long enough half-lives, assist the Penning trap mass spectrometer by removing isobaric contaminants~\cite{ Wienholtz2013, Wolf2012OnlineSO, WOLF2013123,DICKEL2015172,REITER2021165823,CHAUVEAU2016211}, see Sec.~\ref{sec: Expanding Mass Separation}, that may be necessary for a successful Penning trap mass measurement. In summary, an MR-ToF MS at FRIB can greatly expand the reach of FRIB's mass measurement program complementing Penning-trap high-precision mass measurements and enable studies very close to the driplines.  

\subsection{\label{sec: Expanding Mass Separation} Providing isobarically and isomerically purified ion beams at high rates}
\begin{figure}[t]
\centering
\includegraphics[width=\columnwidth]{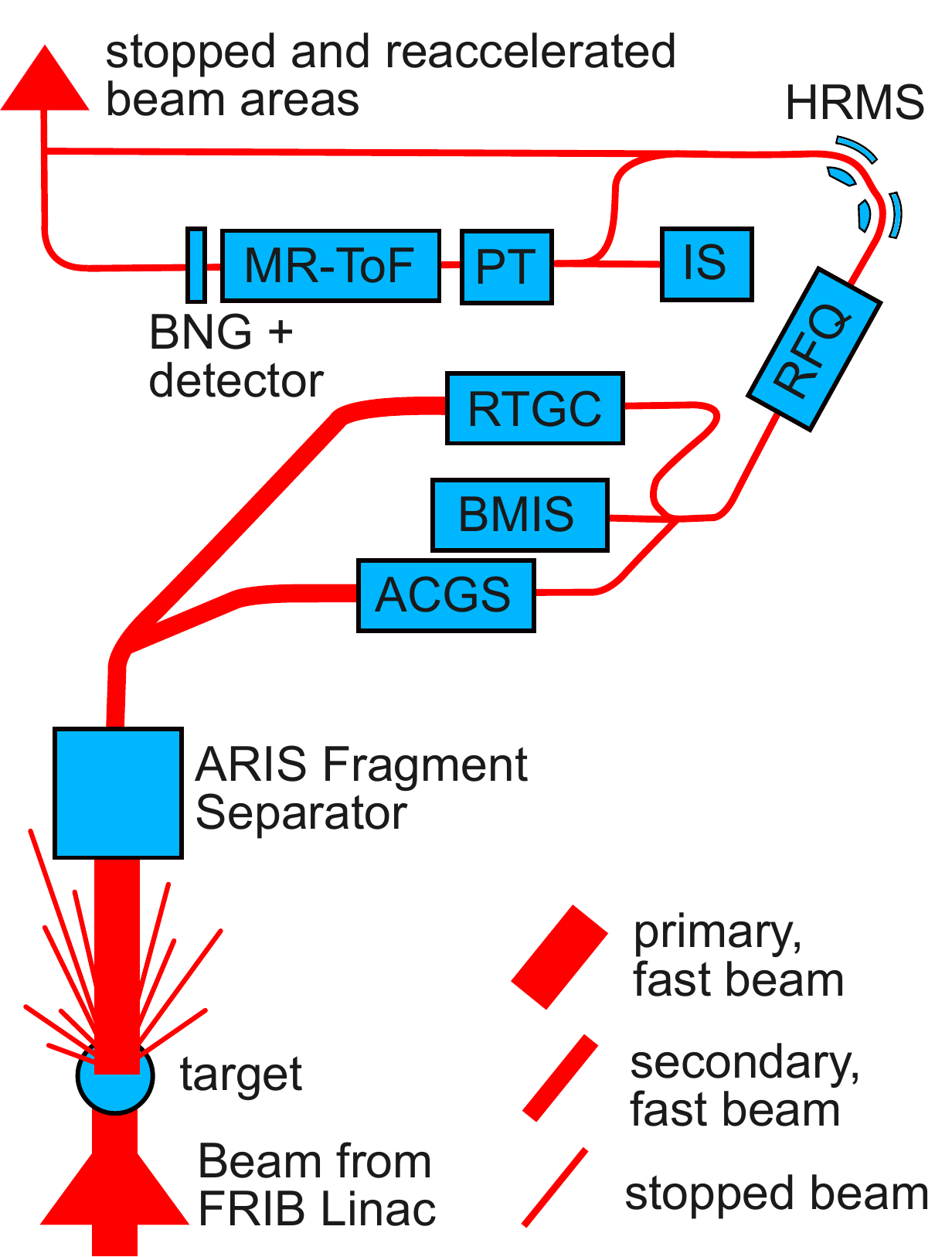}
\caption{Schematic overview of FRIB's isotope production and stopped beam delivery including planned developments. After acceleration in FRIB's superconducting linear accelerator~\cite{PORTILLO2023151}, the primary beam impinges on a target creating a wide variety of different ion species, that are purified with ARIS and stopped in RTGC or ACGS. In the current configuration, RTGC and ACGS each have their own RFQ, while BMIS does not have one. In the future, RTGC, ACGS, and BMIS should share a single RFQ and the present magnetic dipole mass separator should be replaced by an high-resolution magnetic mass separator (HRMS). Furthermore, an MR-ToF MS is planned to be added. It consists of its own offline ion source (IS) for beam tuning and calibration, a dedicated Paul-trap cooler buncher (PT), the MR-ToF device, a Bradbury-Nielsen gate (BNG) and a retractable ion detector.}
\label{fig:entire setup}
\end{figure}

\textbf{Isobarically pure beams --}
 The purity and rate of the delivered rare-isotope beam are critical parameters for FRIB’s science program with stopped and reaccelerated beams~\cite{VILLARI2023350}. Fig.~\ref{fig:entire setup} shows a schematic of FRIB's isotope production and stopped beam delivery process. 
  High-energy, rare-isotope beams produced via projectile fragmentation at FRIB are (partly) purified by the Advanced Rare Isotope Separator (ARIS)~\cite{PORTILLO2023151} and stopped using one of FRIB’s two gas stoppers, either a linear Room Temperature Gas Catcher (RTGC)~\cite{SUMITHRARACHCHI2020305} or the Advanced Cryogenic Gas Stopper (ACGS)~\cite{LUND2020378}. 
Isobaric beam contaminants are commonly present in the beams extracted from the gas stoppers. These can be stable molecules of impurities in the stopping gas, which can have orders of magnitude higher rates than the rare isotope of interest, radiomolecules formed by reactions of stopped ions with gas impurities, or decay daughters of the stopped, rare-isotope ions. Furthermore, isobaric contamination can also be present in the beam incoming to the gas stopper from ARIS~\cite{PORTILLO2023151}. Beams produced by FRIB's Batch Mode Ion Source (BMIS)~\cite{SUMITHRARACHCHI2023301}, which is designed to deliver stable or long-lived ion species e.g. generated by FRIB's isotope harvesting program~\cite{Abel_2019}, are typically also subject to isobaric contamination. Since the quantity of the isotope of interest introduced into the ion source is often very small ($\approx 100$~ng), even minimal contamination from materials within the ion source itself could lead to significant beam impurity.

In the current configuration of the stopped beam facility, beams extracted from the gas catchers pass through a radiofrequency quadrupole (RFQ) which acts as a beam cooler and differential pumping barrier, and are further purified by a magnetic dipole mass separator with a mass separation power of $\approx 1\,500$. Beams produced by BMIS are passed directly to the mass separator without undergoing any cooling by an RFQ. The mass separation power of the dipole magnet is therefore reduced to $\approx$ 300. Afterward, the beam is delivered to the desired end station within the stopped beam facility or sent to FRIB's ReAccelerator (ReA)~\cite{VILLARI2023350,Villari:IPAC2016-TUPMR024}. 
The mass separation power of the dipole magnet is often not high enough to remove all isobaric contamination, whereas an MR-ToF MS would, for most cases, provide isobarically purified beams to downstream experiments within FRIB's stopped and reaccelerated beam facility. 

\textbf{Isomerically pure beams --}
Decay spectroscopy measurements often require beams consisting of only one isomeric state at the highest possible rate.
As an example, the LEBIT Penning trap~\cite{RINGLE201387} has recently been used to successfully separate three isomeric states in $^{70}$Cu for total absorption spectroscopy with SuN++~\cite{SIMON201316, RONNING2026170930, Ronning2025PLB}. However, due to space-charge effects in the Penning trap, only 10 ions/s could be purified, whereas 10,000 ions/s would have been available. 
 Additionally, the Penning trap cannot be used if the ratio of ions of interest to contaminants is too high. Hence, FRIB would benefit from an isobar and isomer separator capable of providing purified beams at high rates to all experimental stations in the stopped and reaccelerated beam facility. 

This isomer separator would, among others, enable the discovery and mass determination of many new isomeric states, see e.g.~\cite{Dickel2024}, as well as new decay spectroscopic measurements for investigating shape coexistence~\cite{RevModPhys.83.1467}, for constraining neutron-capture reaction rates~\cite{PhysRevLett.113.232502, PhysRevLett.116.242502}, and for measuring the $\beta$-decay intensity~\cite{PhysRevLett.117.142701}. The decay spectroscopic data also serve as a benchmark for theoretical models that provide large-scale calculations for the r-process~\cite{PhysRevLett.117.142701}. Understanding the population and depopulation of isomeric states is critical to determining their impact on the overall abundances of nuclei produced in e.g. neutron-star mergers or supernovae~\cite{Misch2024}. One mechanism available for depopulating the isomeric state is Nuclear Excitation by Electron Capture (NEEC)~\cite{GOLDANSKII1976393,Chiara2018}, which could also enable the utilization of the large amounts of energy stored in a long-lived isomeric state~\cite{Carroll2024}. An isomerically pure beam with high intensity is of utmost importance for all of these studies, but with FRIB's current techniques not reachable. 


Current state-of-the-art MR-ToF MSs routinely achieve mass resolving powers $m/\Delta m >5\cdot 10^5$ within a few milliseconds of storage time, allowing the resolution of almost all (molecular) isobars as well as $\approx$ 50\% of all known isomers with half-lives above 10~ms~\cite{Dickel2024}. If a mass resolving power of $3\cdot 10^6$ could be reached, 90\% of all known isomers with half-lives above 10~ms could be resolved in time-of-flight~\cite{Dickel2024}.\footnote{Note that the device's mass resolving power $m/\Delta m$ describes the capability of the device to distinguish two different ion species with masses $m_1$ and $m_2$. For a full mass separation of the stored ion species as required to fully isolate and transmit only the ions of interest to subsequent experimental stations, the device's mass separation power needs to be some factor $\beta$ larger than the $m/\Delta m$ ratio of the stored ion species. We hence distinguish between mass resolving and mass separation power throughout the manuscript. The former refers to the ability to distinguish between two ion species with similar mass, while the latter describes the ability to fully isolate and transmit only the desired ion species.} Space-charge effects, however, pose a challenge for MR-ToF mass separation in cases where excessively many ions are confined in the MR-ToF device~\cite{BOLOTSKIKH2011146, RosenbuschAIP2013, RosenbuschAIP2015, PhysRevE.104.065202, MAIER2023168545}. Current state-of-the-art MR-ToF MSs that store ions at an ion beam energy of around 2~keV can achieve mass separation capabilities of $m_1/(m_2-m_1)>10^5$ for ion rates of up to 2000 ions/s for two simultaneously stored ion species with masses $m_1$ and $m_2$. 
Simulations (see Ref.~\cite{MAIER2023168545} and Sec.~\ref{sec: Ion Flux} in this work) predict that up to $10^5$ ions/s can be separated for $m_1/(m_2-m_1)>10^5$ when increasing the energy of the stored ions to 30~keV and improving the geometries of the MR-ToF device.

\subsection{\label{sec: Beam diagnostics} Beam composition diagnostics}
MR-ToF MSs enable the simultaneous identification of multiple isobars within a single measurement, allowing for quick diagnostics of the different ion species present in the ion beam, together with their relative intensities~\cite{REITER2020431, AU2023375, PhysRevC.107.064604}. This capability makes the MR-ToF MS particularly valuable for diagnosing the composition of ion beams delivered from the gas stoppers~\cite{SUMITHRARACHCHI2020305,
LUND2020378} or BMIS~\cite{SUMITHRARACHCHI2023301}, and for the corresponding optimizations of the ions of interest. Additionally, MR-ToF MSs allow for unambiguous ion identification—also referred to as tagging—which is essential for verifying and calibrating particle identification (PID) systems~\cite{HORNUNG2023257}. At in-flight facilities like FRIB, such systems typically rely on time-of-flight, energy deposition, and magnetic rigidity measurements of the fast beam, but require the unambiguous identification of one ion species to guarantee a correct calibration. In contrast to other tagging methods, mass tagging provides rapid identification and is not constrained by the decay characteristics of the isotopes~\cite{HORNUNG2023257}.
Minimizing the time it takes to develop and characterize the beams prior to delivery for user experiments will make the best use of available beam time and maximize scientific output.

\section{\label{sec:Setup}Proposed experimental setup}
A highly selective and high-flux MR-ToF MS is currently designed at FRIB. A general overview of the entire stopped beam delivery process, including the planned upgrades such as the MR-ToF MS, is shown in Fig.~\ref{fig:entire setup}. In the upgraded delivery process, the beam extracted from the RTGC~\cite{SUMITHRARACHCHI2020305} or ACGS~\cite{LUND2020378} gas stopper, or the beam produced by BMIS~\cite{SUMITHRARACHCHI2023301} is guided to a room-temperature, buffer-gas-filled radiofrequency quadrupole (RFQ)~\cite{BARQUEST201718, Lapierre_2024}, which acts as a beam cooler, producing beams with a 95\%-emittance of a few $\pi$ mm mrad and a few eV energy spread. It can also be used to transfer the continuous beam to bunched beams with a few $\mu$s bunch widths, if needed. The beam is then sent to a High-Resolution Magnetic Mass Separator (HRMS), which is expected to achieve a mass separation power of $\approx 10,000$. If this mass separation power is sufficient, the purified beam can be sent directly to the stopped beam or reaccelerator beam facility. If a higher mass separation power is required, the beam is guided to the planned MR-ToF MS setup. First, the ion beam is injected into a buffer-gas-filled Paul-trap cooler-buncher (PT) that prepares ion bunches with less than 4~ns bunch widths (FWHM) at the position of the MR-ToF device for $^{133}$Cs, as our ion optical simulations show, see Sec.~\ref{sec:Performance}. Second, these ion bunches are mass-separated in the MR-ToF device. Third, the contaminants will be removed with a Bradbury-Nielsen gate (BNG)~\cite{PhysRev.49.388, PLA20084560, WOLF201282} installed downstream of the MR-ToF device (see Sec.~\ref{sec: Removal of contaminants} for more details) and the purified ion beam is sent to the stopped or reaccelerated beam facility. If needed, the in the MR-ToF device mass-separated ions can also be guided back into the Paul-trap cooler-buncher for mass-selective retrapping~\cite{doi:10.1007/s13361-017-1617-z}. An ion source (IS) located directly upstream of the Paul trap (see Fig.~\ref{fig:entire setup}) is used for offline tests and calibration. 

\begin{figure*}[t]
\centering
\includegraphics[width=1\textwidth, angle=0]{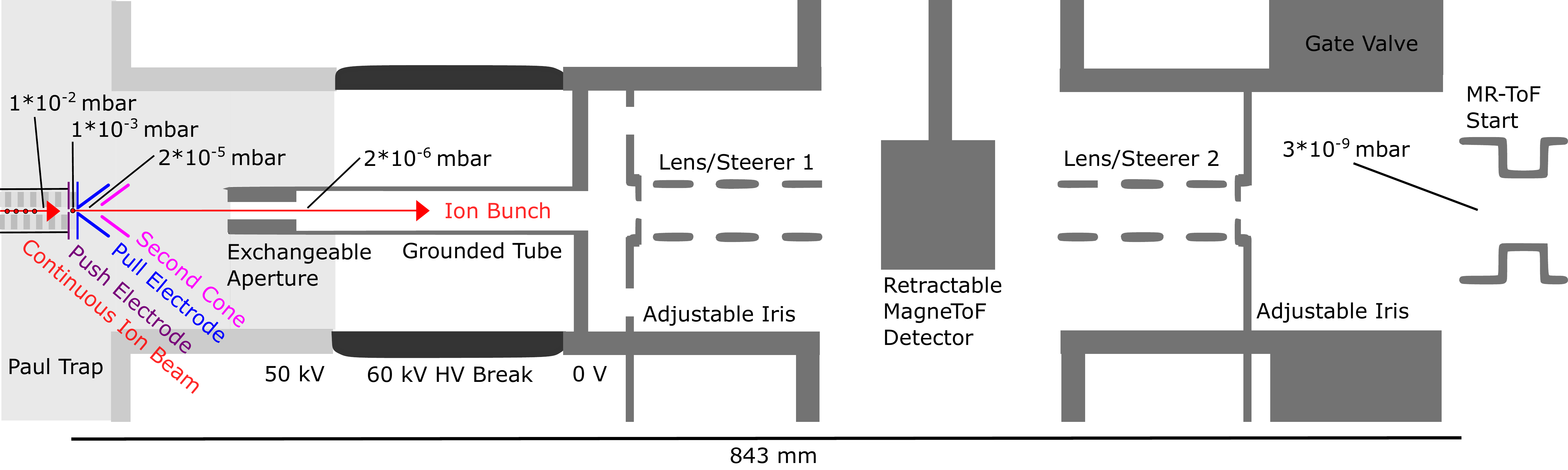}
\caption{Cross section view of the downstream side of the Paul-trap cooler buncher and the injection optics into the MR-ToF device. Helium buffer gas is injected into the first cooling stage located upstream of the push electrode in order to have a pressure of $\approx 1\cdot 10^{-2}$ mbar for optimal ion accumulation. The second cooling stage is located between the push (depicted in purple) and pull (depicted in blue) electrodes and the buffer-gas pressure in this region drops to  $\approx 1\cdot 10^{-3}$ mbar. The simulated Helium pressures along the remaining beamline due to the flow of the buffer-gas atoms are indicated. Pressure simulations assume two 600 l/s turbo pumps located below the Paul-trap cooler buncher and below the MagneToF detector.}
\label{fig: injection optics}
\end{figure*}

\subsection{Paul-trap cooler-buncher}
In the detailed MR-ToF setup description provided next, we begin with a brief overview of the ion preparation in the Paul trap cooler-buncher directly upstream of the MR-ToF device, as it largely influences the MR-ToF performance. The Paul trap simulations were based on the design of a cooler-buncher developed and commissioned at the University of Edinburgh, for which a publication is currently in preparation by the Edinburgh group~\cite{PascalsPaultrap}. 

For the proposed setup at FRIB, the Paul trap is floated to 50~kV. It consists of two pressure regions with $10^{-2}$ and $10^{-3}$ mbar, see Fig.~\ref{fig: injection optics}. 
The first region facilitates ion accumulation and initial cooling of the injected continuous or bunched ion beam, whereas the second region forms ion bunches with properties optimized for injection into the MR-ToF device. 
The two pressure regions are separated by an electrode (`push electrode') with an opening hole of 2~mm diameter 
which acts as a differential pumping barrier. The low-pressure region has open sides leading to a pressure drop by a factor 10 compared to the high-pressure region as simulations in Molflow~\cite{Kersevan:IPAC2019-TUPMP037} show. The open sides can, at a later stage of the project, also facilitate transversal laser access for Doppler and sympathetic cooling~\cite{Sels2022}. Due to the lower pressure in the second region, reheating effects during ion extraction from the Paul trap are reduced. 
The axial potential wall in the second region is formed by the push and pull electrodes and the RF segment between. The DC potential of the RF segment is 10 V and the push and pull electrodes are biased to 60 and 62.3~V with respect to the 50~kV high-voltage platform. The effective trapping volume of the second cooling stage is about 3x3x3 mm$^3$ and the axial trap depth is $\approx 10$~V. 

After a few milliseconds of storage in the low-buffer gas region, the push-and-pull electrodes are switched to 300 and -300 V with respect to the high-voltage platform, and the ions are released from the Paul trap. The extraction field strength is $\approx 125$~V/mm, enabling the production of ion bunches with bunch widths of $< 4$~ns (FWHM) at the time focus position matched to the MR-ToF device. The pull electrode has an extraction hole with a diameter of 2 mm and is, for the modified design at FRIB, connected to the first extraction cone. The second cone is biased at -5~kV with respect to the 50~kV bias of the Paul-trap platform. The cones create an additional extraction gradient that is helpful in matching the time focus position with the MR-ToF device~\cite{YAVOR2009283}. Additionally, simulations~\cite{LECHNER2024169471} showed that the double-cone extraction optics could limit reheating effects during ion extraction from the Paul trap. Due to the cones, a high ion acceleration in close proximity to the exit hole of the Paul trap, where the helium pressure is still quite high, is avoided~\cite{LECHNER2024169471}. 

\begin{figure*}[t]
\centering
\includegraphics[width=1\textwidth]{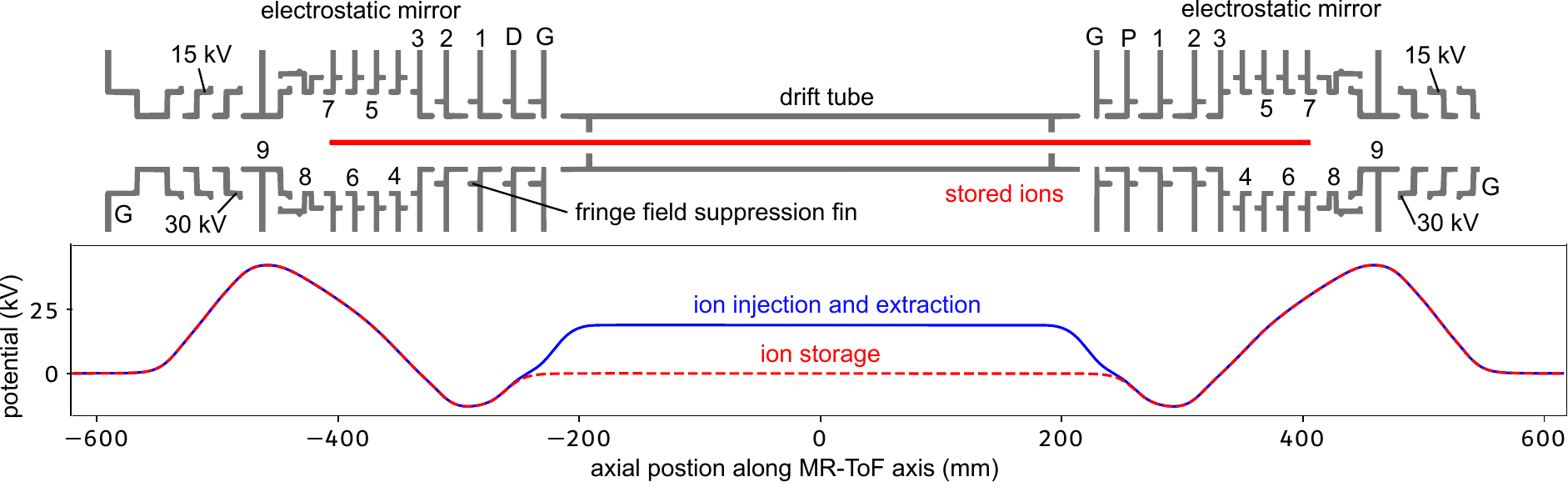}
\caption{Top: Cross section view of FRIB's 30-keV MR-ToF device. The simulated ion trajectories for 300 ions performing 300 revolutions are depicted in red. The mirror electrodes are labelled with 1-9, G stands for a grounded electrode, D is the deflector electrode and P is the pickup electrode, see text for details. Bottom: Electric potential along the central axis of the MR-ToF device for ion storage at 31.094 keV (red dashed) and for ion injection and extraction (blue). }
\label{fig: MR-ToF setup}
\end{figure*}

\subsection{MR-ToF injection optics}
After the end of the second cone of the Paul trap, a grounded tube is installed, which shields a ceramic break separating the 50 kV platform from the remaining beamline at ground potential (see Fig.~\ref{fig: injection optics}). The grounded tube has an entrance aperture with a length of 40 mm and a diameter of 12 mm in order to limit the flow of helium gas from the Paul trap to the MR-ToF device. After the grounded tube, two high-voltage einzel lens/steerer assemblies are positioned that surround a retractable MagneToF detector for the detection of ions (see Fig.~\ref{fig: injection optics}). They are biased to -45~kV and -25~kV, respectively.  Two adjustable iris shutters based on the design in Ref.~\cite{Klink_2024} are placed between the grounded tube and the first einzel lens/steerer assembly, as well as between the second einzel lens/steerer assembly and a gate valve. They enable a restriction of the helium flow towards the MR-ToF device. Pressure simulations in Molflow~\cite{Kersevan:IPAC2019-TUPMP037} show that for 10~mm iris apertures, the on-axis pressure at the entrance of the MR-ToF device is $3\cdot 10^{-9}$ mbar and $7\cdot 10^{-9}$ mbar for 20~mm diameters. Iris apertures down to 3 mm are feasible without observing any ion losses in simulations, which would allow a pressure reduction down to $1\cdot 10^{-9}$ mbar. 
At the entrance of the MR-ToF device, a grounded electrode, a 15~kV electrode, and a 30~kV electrode are located, as shown in Fig.~\ref{fig: MR-ToF setup}. These three electrodes aid the transversal focusing of the ion bunch into the MR-ToF device. 

\subsection{MR-ToF device}
The MR-ToF device, also shown in Fig.~\ref{fig: MR-ToF setup}, advances the design of the high-voltage device of the MIRACLS collaboration at ISOLDE/CERN~\cite{MAIER2023167927, MAIER2023168545, MAIER2025170365}. In order to keep all electric field gradients present in the entire setup for ion storage at 30~keV beam energy below 3.6~kV/mm (a value often referred to as 'safe operational limit'), the number of mirror electrodes has been increased from 6 at MIRACLS to 9 for FRIB. The increased number of mirror electrodes also results in improved performance in the ion-optical simulations (see Sec.~\ref{sec:Performance}). This is likely due to the greater number of parameters available for optimally shaping the electric fields to align with the relationships discussed in Ref.~\cite{MAIER2025170365}. The potentials applied to the mirror electrodes are stated in Tab.~\ref{tab:MRToFSetting} in the appendix. Fig.~\ref{fig: MR-ToF setup} shows the potential distribution along the central axis of the MR-ToF device for ion injection/extraction, and for ion storage. 

The ion beam energy in front of the MR-ToF device is 50~keV and the ions are captured at 30~keV beam energy in the MR-ToF device via the technique of in-trap lift switching~\cite{WOLF20128}. To this end, the drift tube is switched from 18915~V to ground once the ion bunch is in the center of the drift tube. 
Between the drift tube and the mirrors, a deflection/pickup electrode and a grounded electrode are placed. The deflection (D) and pickup (P) electrodes are azimuthally segmented into four pieces so that they can act as steering electrodes during ion injection. The deflection electrode can additionally be used for the in-situ removal of the stored ion species~\cite{Fischer20182} (see also Sec.~\ref{sec: Removal of contaminants}) while the pickup electrode can be used for in-situ ion monitoring and diagnostics by the pickup of image-charge signals~\cite{doi:10.1021/acs.analchem.7b02797} if it is not required for steering during ion injection. The grounded electrode partially shields the pickup electrode from electrical noise induced by the switching of the central drift tube but is mainly required for mounting purposes. 

The three innermost mirror electrodes (1-3)  facilitate radial refocusing of the ion bunch, whereas electrodes~4-9 axially confine the ions. A radius of 40~mm was chosen for the six outermost mirror electrodes 4-9. This radius is large enough to facilitate a smooth potential distribution in the region of the ions, even for 8~mm gaps between the neighboring electrodes that are required to keep electric field strengths well below 3.6~kV/mm in the entire apparatus. 
In order to avoid large potential differences between mirror electrodes 3 and 4, the radius of mirror electrodes 1-3 is reduced to 20 mm.
The radius of the outermost mirror electrode~9 is at its outer end also reduced to 20 mm to shield the MR-ToF region from external fields. Also, the grounded, 15~kV, and 30~kV electrodes have a 20 mm inner radius. Each electrode is equipped with fringe field suppression fins that ensure a smooth potential distribution between neighboring mirror electrodes. 

Downstream of the MR-ToF device, a BNG is installed, which deflects the mass-separated contaminants. More details on contaminant removal are provided in Sect.~\ref{sec: Removal of contaminants}. A retractable MagneToF detector, located downstream of the BNG, is used for beam tuning, diagnostics, and high-precision mass measurements. If the MR-ToF MS is operated in mass separation mode, the MagneToF detector is retracted, and the purified ion beam is sent to the stopped and reaccelerated beam facility.

\section{\label{sec:Performance}Simulated performance}
Various ion-optical simulations, from the buffer-gas cooling process in the Paul trap to the extraction of the ions from the MR-ToF device, were performed in SimIon~\cite{SimIon} and 3DCylPIC~\cite{RINGLE201142, PhysRevE.104.065202}. Applied potentials and geometries were optimized for the highest mass resolving power and ion flux possible, following the simulation and optimization procedure outlined in Refs.~\cite{MAIER2025170365, MAIER2023168545}. It was ensured that the slopes of the two linear sections of the MR-ToF mirrors, the refocusing lens section with its parabolic potential distribution, the length of the field-free region within the drift tube, and the energy of the stored ions are all well matched, such that the highest possible mass resolving power can be achieved. The simulations were performed with $^{133}$Cs$^+$ ions. Individual results were cross-checked for $^{24}$Mg$^+$ ions and for ions with a mass of 250~u. An overview of the simulated performance of the optimized MR-ToF setup, as presented in Sec.~\ref{sec:Setup}, is provided next.

\subsection{\label{sec: Mass Resolving Power} Mass resolving power}
The mass resolving power of an MR-ToF MS can be expressed as~\cite{WOLF2013123}
\begin{equation}
R= \frac{m}{\Delta m} = \frac{t}{2\Delta t}= \frac{t_0+rt_1+t_d}{2\sqrt{\Delta t_0^2+(r\Delta t_1)^2}}.\label{R formula}
\end{equation}
The total time of flight $t$ refers to the duration between the extraction of ions from the Paul trap and their arrival at the detector downstream of the MR-ToF device, while the temporal spread (full-width-at-half-maximum FWHM) of the ion bunch upon detection is denoted by $\Delta t$. The flight time $t$ consists of three distinct components: the initial transport time $t_0$ from the Paul trap to the central plane of the MR-ToF device, the storage time $t_s = r t_1$ within the MR-ToF device, where $r$ is the number of revolutions and $t_1$ is the duration of a single revolution, and the final transport time $t_d$ from the MR-ToF center to the detector following ejection. When the ion bunch passes the central plane of the MR-ToF device for the first time, it has an initial bunch width $\Delta t_0$. $\Delta t_1$ represents the incremental bunch width broadening introduced per revolution within the MR-ToF device.

\begin{figure}[t]
\centering
\includegraphics[width=1\columnwidth]{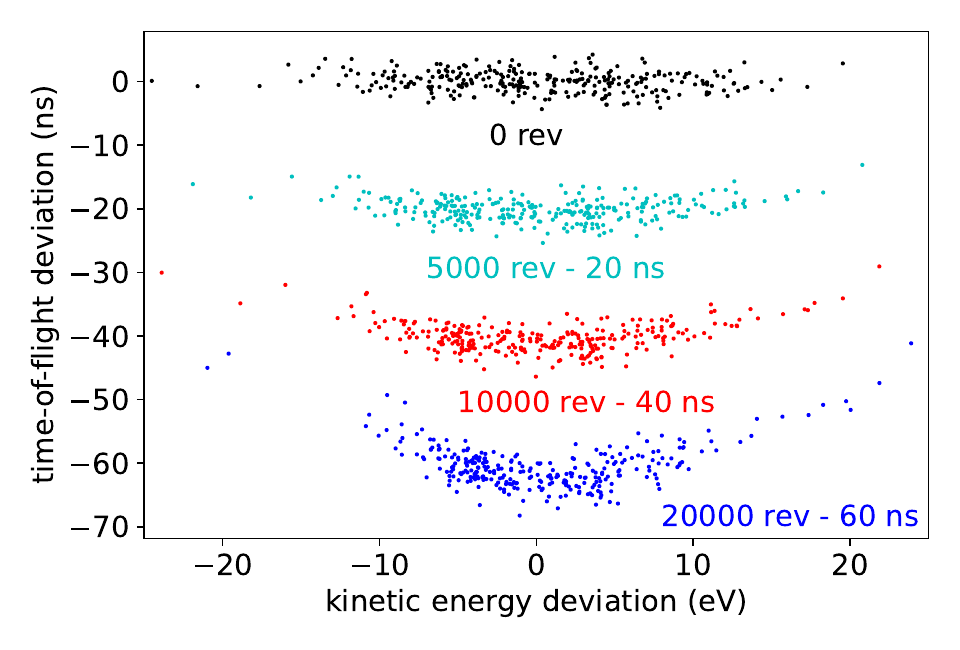}
\includegraphics[width=1\columnwidth]{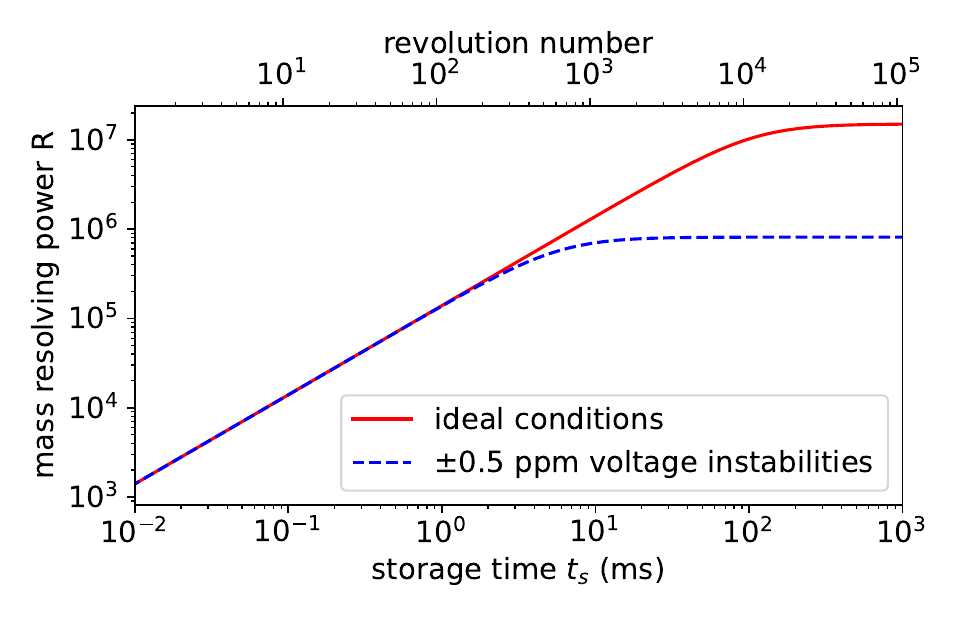}
\caption{(Top) Deviations in the time-of-flight centroid as a function of the deviation in the kinetic energy centroid for 300 simulated ions and various different revolutions in ideal conditions. (Bottom) Simulated mass resolving power as function of storage time or revolution number for simulations in ideal settings as well as when adding $\pm 0.5$~ppm voltage instabilities to the mirror electrodes.}
\label{fig: R sim}
\end{figure}

Fig.~\ref{fig: R sim} shows the longitudinal phase space diagram for various revolution numbers in the MR-ToF device, as well as the simulated mass resolving power as a function of storage time and revolution number. 
After an initial linear increase, the mass resolving power levels off at $R_{\mathrm{inf}}=t_1/(2\cdot \Delta t_1)\approx 2\cdot 10^7$. The initial time spread $\Delta t_0$ is primarily responsible for determining the storage time required to achieve a given mass resolving power $R<R_{\mathrm{inf}}$. In order to achieve a small $\Delta t_0$, the time focus needs to be matched to the MR-ToF device's middle plane. The distance between the Paul trap and the MR-ToF device, as well as the Paul trap extraction settings, define the size of $\Delta t_0$ for which the time focus can be matched to the detector plane. $\Delta t_0$ can be as small as 3.7~ns according to our simulations employing static mirror potentials. Bunch widths shorter than 3.7~ns are expected to be possible if the dynamical time focus shift technique~\cite{DICKEL20171} is applied, which, however, requires a switching of the high-voltage mirror electrodes. Due to the increased technical challenge, this is not foreseen during the first years of operation of the MR-ToF MS. For 3.7~ns time spreads, a mass resolving power of $1\cdot10^5$, $1\cdot10^6$ or $5\cdot10^6$ can be reached within 0.74, 7.4 or 38.5~ms of storage time, respectively.  An initial time spread of 3.7 ns corresponds to an energy spread of 18~eV, which is significantly smaller than the energy spread tolerance of the MR-ToF MS. Even for artificially increased energy spreads of 100~eV, the simulated $R_{\mathrm{inf}}$ is larger than $4\cdot10^6$. 

\begin{figure}[t]
\centering
\includegraphics[width=\columnwidth]{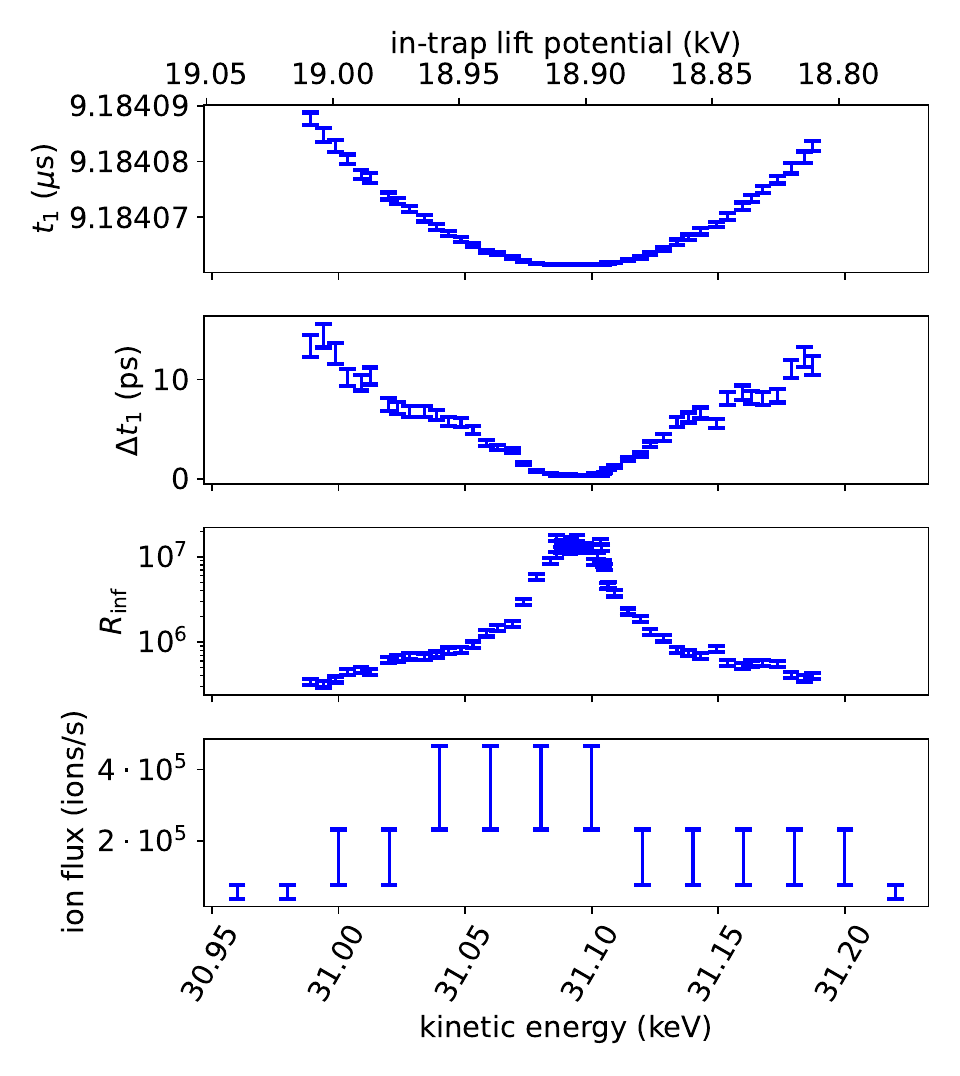}
\caption{Revolution period $t_1$, ion-bunch broadening per revolution $\Delta t_1$, maximally attainable mass resolving power $R_\mathrm{inf}$ and ion flux as function of the in-trap lift potential or the kinetic energy of the stored $^{133}$Cs$^+$ ions. The ion flux simulations were done with $m_1/(m_2-m_1)=10^5$ and an abundance ratio of 1 employing the injection optics that leads to the highest mass resolving power.}
\label{fig: t1_dt1_Rinf_flux_vs_E}
\end{figure}

Fig.~\ref{fig: t1_dt1_Rinf_flux_vs_E} shows the time for one revolution $t_1$, the peak-width broadening per revolution $\Delta t_1$, the mass resolving power $R_\mathrm{inf}$ and the ion flux for two simultaneously stored species with masses $m_1=133$~u and $m_2=133.00133$~u (see section~\ref{sec: Ion Flux} for more details) as functions of the in-trap lift voltage. For in-trap lift potentials from 18895 to 18915 V $R_\mathrm{inf}$ is above $1\cdot10^7$. For in-trap lift potentials from 18870 to 18950 $R_\mathrm{inf}$ is still above $1\cdot10^6$. For these in-trap lift potentials the MR-ToF device is operated in isochronous mode, where $t_1$ is approximately constant and the mass resolving power and ion flux are the highest. Despite the setup exhibiting only second-order energy focusing, as evidenced by the quadratic dependence of the revolution period as function of in-trap lift potential (see Fig.~\ref{fig: t1_dt1_Rinf_flux_vs_E} (Top)) as well as the quadratic dependence of the time-of-flight deviations on the kinetic energy deviations shown in Fig.~\ref{fig: R sim} (Top), $\Delta t_1$ is $\approx 0.25$~ps for the typical operational parameters and $R_{\mathrm{inf}} \approx 2\cdot 10^7$. Notably, numerous potential configurations were found that provide third-order energy focusing, enabling $R_{\mathrm{inf}} > 1\cdot 10^6$ for in-trap lift potentials spanning more than a 400~V range. However, even with computing times hundred times longer than those required to find optimal second-order focusing conditions with $R_{\mathrm{inf}} > 1\cdot 10^7$, these third-order configurations did not yield mass resolving powers exceeding $3\cdot10^6$ in any of the 20 different MR-ToF geometries tested during this work.

Due to power supply instabilities, temperature fluctuations, residual gas collisions, and space-charge effects (see Sec.~\ref{sec: Ion Flux}) the mass resolving power will be significantly reduced. When accounting for peak-to-peak voltage instabilities of $\pm 0.5$~ppm for all mirror electrodes in the simulations and when ignoring any possible cancellation effects, the mass resolving power drops to $R_{\mathrm{inf}}=8\cdot 10^5$.  The external stabilization of 60 kV FUG power supplies with less than 0.5~ppm peak-to-peak voltage fluctuations has recently been successfully demonstrated~\cite{10.1063/5.0218649, PASSON2025101818}. More details regarding the impact of non-ideal conditions and possible mitigation strategies will be provided in an upcoming manuscript.

\subsection{\label{sec: Ion Flux} Ion flux}
Space-charge effects, primarily due to ion-ion interactions, can severely limit the mass resolving power in an MR-ToF setup~\cite{BOLOTSKIKH2011146, RosenbuschAIP2013, RosenbuschAIP2015, PhysRevE.104.065202, MAIER2023168545}.  To assess the impact, we perform simulations that take space-charge effects in the Paul trap and MR-ToF device into account. The space-charge simulations in the Paul trap are performed using SimIon's space-charge repulsion-factor method, in analogy to Ref.~\cite{LECHNER2024169471}. These simulations show that the longitudinal and transversal emittances of the extracted ions are increased by a factor of 2-3 when 100,000 ions are simultaneously stored in the Paul trap, compared to the case where fewer than 10,000 ions are simultaneously stored. The extracted ions are then injected into the MR-ToF device. To inspect the impact of space-charge effects on the MR-TOF's separation capabilities, we simulate ion bunches containing two isobaric species of masses $m_1$ and $m_2$. Both ion species are present in equal amounts, corresponding to an abundance ratio of 1. As shown in Ref.~\cite{MAIER2023168545} the ion flux does not depend on the abundance ratio for all the simulated abundance ratios between 0.02 and 50. 
The MR-ToF space-charge simulations are performed in SimIon in analogy to Ref.~\cite{MAIER2023168545}, but also in 3DCylPIC~\cite{RINGLE201142, PhysRevE.104.065202}. The latter implements space-charge effects more realistically via a particle-in-cell approach and also takes image charges into account. The maximal ion flux $\Phi$ for a given $m_1/(m_2-m_1)$ ratio is defined as the highest number of ions $N_\mathrm{max}$ that can be mass separated per unit time
\begin{equation}
    \Phi =\frac{N_\mathrm{max}}{t_\mathrm{proc}}. 
\end{equation}
The processing time $t_\mathrm{proc}$ is determined by the longer of the two durations: the ion preparation time $t_{\mathrm{prep}}$=5~ms in the Paul trap and the MR-ToF storage time $t_s$ required for mass separation. This approach assumes that the preparation of ions in the Paul trap happens while another ion bunch is mass separated in the MR-ToF device.

The maximum number of ions (from both ion species combined) that can be simultaneously stored in the MR-ToF device without causing more than 7\% peak overlap due to space-charge effects is given by $N_\mathrm{max}$.  Hence, as long as the total number of stored ions remains below $N_\mathrm{max}$, the two ion species can still be successfully mass separated with less than 7\% contamination from the respective other ion species.

\begin{figure}[t]
\centering
\includegraphics[width=\columnwidth]{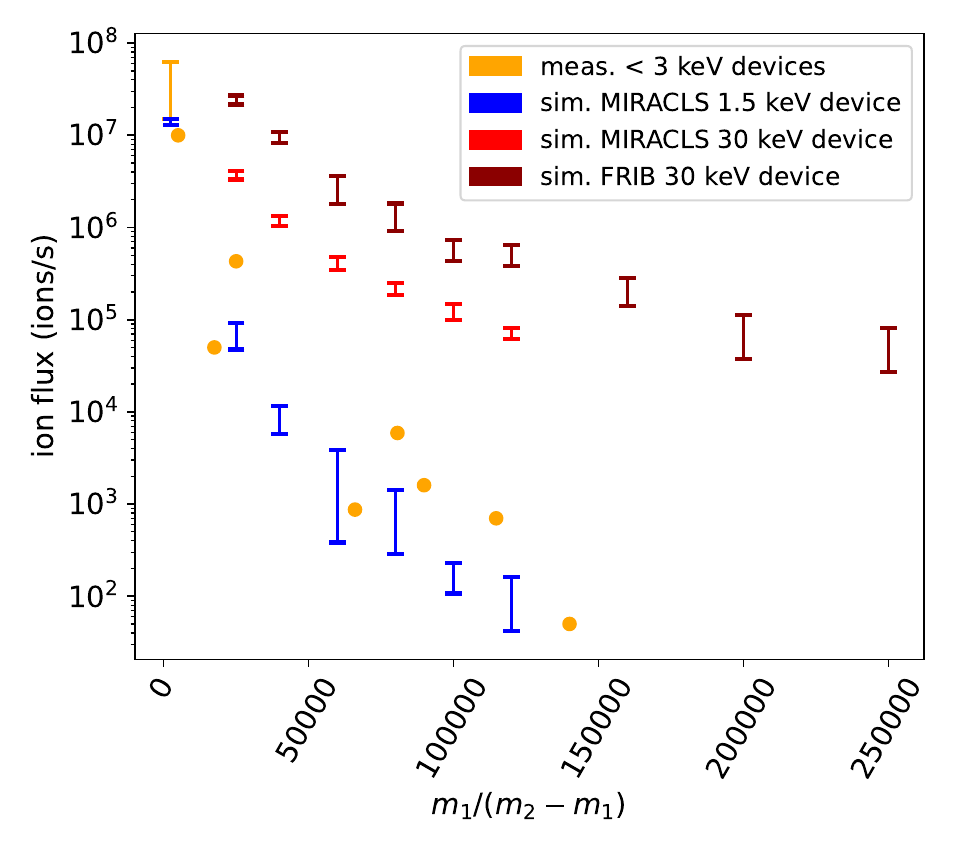}
\caption{Maximal ion flux versus $m_1/(m_2-m_1)$ for FRIB's 30 keV MR-ToF MS (simulation) in comparison to MIRACLS' 30 keV and 1.5 keV MR-ToF devices (simulation) and different low-energy MR-ToF MSs (experiment). The simulations of FRIB's 30 keV MR-ToF MS are performed for $A=133$, the ones for MIRACLS' 30 keV device for $A=132$~\cite{MAIER2023168545} and the ones for MIRACLS 1.5 keV device for $A=24$~\cite{MAIER2023168545}. The ion flux measurements for MIRACLS' 1.5 keV device are for $A=28$~\cite{RosenbuschAIP2013, RosenbuschAIP2015}, from ISOLTRAP for $A=152$~\cite{Wienholtzcom} and $A=167, 186, 225$~\cite{BouwmanSummerStudent}, from Giessen for $A=78$~\cite{DICKEL2015172}, and from TITAN for $A=28, 60, 131$ and 151~\cite{REITER2021165823,PhysRevC.103.025811, PhysRevC.104.065803}.}
\label{fig: IOn flux}
\end{figure}

The space-charge simulations in SimIon and 3DCylPIC yield consistent results, supporting the conclusion from Ref.~\cite{MAIER2023168545} that SimIon is a reliable tool for predicting the onset of space-charge effects. Due to the increased beam energy and improved geometry, the simulated maximal ion flux is, on average, two orders of magnitude higher than that of existing state-of-the-art MR-ToF MSs, see Fig.~\ref{fig: IOn flux}. It also exceeds the simulated maximal ion flux of MIRACLS' high-voltage MR-ToF MS with an assumed beam energy of 30 keV by approximately a factor of six. This additional increase is partly due to a different tuning mode of the MR-ToF injection optics at FRIB, which avoids a transversal focus of the ions in close proximity to the turnaround point, at which the ions change their longitudinal direction. In this way, the ions are further spatially spread apart in the mirrors, leading to reduced space-charge effects and an increase in ion flux. When utilizing this tuning mode in FRIB's 30 keV MR-ToF MS, the ion flux values are a factor $\approx 3$ larger than without this tuning mode.
However, this different tuning mode comes at the cost of a reduction in mass resolving power. While it is still possible to achieve a simulated mass resolving power $R_\mathrm{inf}> 10^6$ in FRIB's MR-ToF setup when employing this tuning mode, a suitable mass resolving power could not be reached in the simulations of MIRACLS 30 keV MR-ToF MS.  

In summary, mass separation capabilities of $m_1/(m_2-m_1)>10^5$ are expected to be achievable for ion rates of up to $10^5$ ions/s for two simultaneously stored ion species with masses $m_1$ and $m_2$. This capability enables the delivery of isobarically and isomerically purified ion beams at high intensities to the stopped and reaccelerated beam facility, even for applications that demand high mass separation powers significantly exceeding the expected performance of the High-Resolution Magnetic Mass Separator HRMS.

\subsection{\label{sec: Mass Range} Unambiguous mass range}
After several revolutions, lighter ions may overtake heavier ones due to their increasing time-of-flight separation, leading to ambiguities in determining the number of completed revolutions. To ensure unambiguous mass assignment, only ions within a specific mass-to-charge ($m/q$) range should be considered - specifically, those that complete the same number of revolutions $r$. The condition for this unambiguous mass range is given by~\cite{YAVOR20151}
\begin{equation} \label{eq: turn-number}
    \frac{(m/q)_{\mathrm{max}}}{(m/q)_{\mathrm{min}}} = \left( \frac{r+1}{r}\right)^2.
\end{equation}
According to Equ.~\ref{eq: turn-number} if two ion species with $m_1/(m_2-m_1) =10^6$ or $10^5$ are simultaneously stored in the device, there will be no ambiguity in the revolution number for a third simultaneously stored ion species with a mass $m_3$ as long as $m_1/(m_3-m_1)>10^3$ or $10^2$ for the MR-ToF settings discussed above. The HRMS installed prior to the MR-ToF MS is expected to have a mass separation power $\approx 10^4$. Hence, ambiguity in the revolution number will not be an issue for the present MR-ToF setup. If the mass separation power of HRMS is lower than expected, the deflector electrodes can be used to remove contaminants that are far away in mass before turn-number ambiguity occurs (see Sec.~\ref{sec: Removal of contaminants}).

\subsection{\label{sec: Removal of contaminants} Removal of contaminants}
After a successful mass separation using the MR-ToF MS, contaminant ions, separated by their time-of-flight from the ions of interest, must be removed in order to provide an isobarically or isomerically purified ion beam to subsequent experiments. Various methods exist for current state-of-the-art MR-ToF MSs storing ions at $< 3$ keV beam energy, but their performance for high-voltage MR-ToF MSs is unexplored. In the following, we evaluate the performance of these methods in the context of FRIB’s high-voltage MR-ToF MS and outline the necessary adaptations to ensure efficient contaminant removal without affecting the ions of interest.

\subsubsection{Mass-selective ion ejection technique}  In this approach, the in-trap lift is triggered when the ions of interest are within the drift tube, while contaminants have either not yet entered the drift tube or have already passed through it~\cite{WIENHOLTZ2017285}. Thus, only the ions of interest gain enough energy to exit the MR-ToF device, provided they are sufficiently separated in time-of-flight from the contaminants. 
To leave the ions of interest fully undisturbed while removing the contaminants (for more details and explanations see the appendix), the contaminants need to be $\approx 530$~ns separated from the ions of interest. This results in a time resolution of $\tau_{\mathrm{res}}\approx530$~ns, which means that the $m_1/(m_2-m_1)$ ratio needs to be $\approx$ 130 times smaller than the mass resolving power $R$ at the given storage time for ion bunch widths of 4~ns (FWHM). While the mass-selective ion ejection technique does not require any further equipment, it either severely reduces the mass separation power of the MR-ToF system or requires additional optimization time for the transfer of the ions of interest to subsequent experiments since their ion beam energy is changed by several keV (see appendix for details).

\begin{figure*}[t]
\centering
\includegraphics[width=1\textwidth]{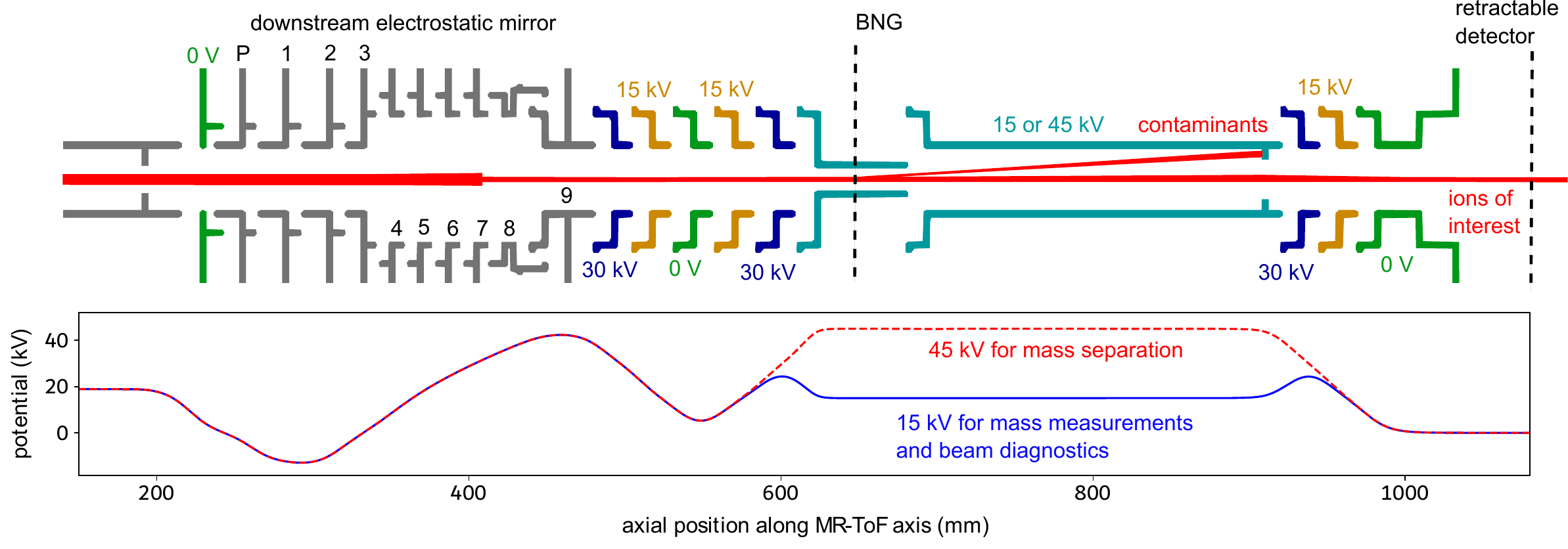}
\caption{Top: Cross section view of the extraction optics of FRIB's 30 keV MR-ToF device. The simulated ion trajectories for 300 ions performing 100 revolutions and being extracted with and without activation of the BNG are depicted in red. The deflection angle is 3.5 degrees. Bottom: Electric potential along the central axis for mass separation (dashed red) and mass measurements and diagnostics (blue).}
\label{fig: BNG Setup}
\end{figure*}

\subsubsection{Bradbury-Nielsen Gates (BNGs)}
Another tool that is widely used for the removal of time-of-flight separated contaminants is a Bradbury-Nielsen Gate (BNG)~\cite{PhysRev.49.388, PLA20084560, WOLF201282}. It is installed directly downstream of the MR-ToF device. A BNG consists of equally spaced parallel wires arranged perpendicular to the beam axis, with alternating potentials applied to adjacent wires. To allow the ions of interest to pass through undisturbed, all wires are set to 0 V when they are passing. When unwanted contaminants are expected, a potential difference is applied between neighboring wires, deflecting the contaminants such that they strike surrounding electrode structures and are consequently removed. To achieve more than 90\% transmission for the ions of interest, a wire diameter of $50~\mu$m and a wire spacing of 0.5~mm were chosen. 

At the unprecedented beam energy of 50 keV downstream of FRIB's 30 keV MR-ToF device, a typical BNG operation would not provide sufficient deflection to remove contaminants. Two potential solutions exist: increasing the voltage applied to the wires, or decreasing the beam energy at the position of the BNG. The former would require too high voltage differences - in dedicated tests, we observed sparks occurring between two wires with 50~$\mu$m diameter that were 0.5~mm apart once the applied potential difference exceeded 2.6 kV in an ultra high vacuum environment. 
The latter solution is feasible: reducing the beam energy from 50 keV to 5 keV was shown to promote enough deflection at safe voltage limits. To accomplish this, a dedicated extraction optics system was designed (see Fig.~\ref{fig: BNG Setup}). After the last MR-ToF mirror electrode (\#9), three electrodes biased to 30, 15 and 0~kV are positioned to maintain consistent electric fields between the upstream and the downstream mirror. Afterwards, the beam passes through electrodes biased to 15 and 30 kV, before the potential is raised to 45 kV at the location of the BNG. A drift electrode, also at 45 kV, is placed downstream, followed by three more electrodes biased to 30, 15 and 0 kV, respectively, which are required for mounting purposes. The downstream end of the drift electrode features a reduced inner diameter of 26 mm, which helps to block deflected contaminants. This electrode configuration reduces the ion beam energy at the location of the BNG to 5~keV. Further energy reduction is possible by raising the bias voltage applied to the BNG and drift electrode above 45~kV. Alternatively, the platform voltage of the Paul trap can be lowered, thereby reducing the overall injection and extraction beam energy and hence also the energy of the ions at the position of the BNG. As long as the Paul trap floating voltage exceeds 46.5~kV, the ions can be stored at a beam energy of 30 keV in the MR-ToF device with a simulated mass resolving power above $10^7$. With $\pm 500$~V applied to the BNG wires, the deflection angle increases to 3.5~degree at 5~keV beam energy and to 5.9~degree at 3~keV. As illustrated in Fig.~\ref{fig: BNG Setup}, a deflection angle of 3.5 degrees is sufficient to remove the contaminants. 

The achievable time resolution of the BNG can be approximated by the time required for an ion to traverse a distance equivalent to twice the wire spacing~\cite{Yoon2007},
\begin{equation}
    \tau_{\mathrm{res}}= 2d \sqrt{\frac{m}{2E}},
\end{equation}
since one wire spacing $d$ away from the plane of the BNG the electric field has almost vanished. For a kinetic energy $E=5$~keV and $d=0.5$~mm, $\tau_{\mathrm{res}}$ is 5, 12 and 16 ns for ions with mass $m$ of 24, 133 and 250~u, respectively. The time constants of high-voltage switches to go from a transparent to a non-transparent state are of the same order. To effectively suppress contaminants while allowing the desired ions to pass undisturbed, the MR-ToF MS must offer a mass resolving power about 5-10 times greater than the $m_1/(m_2-m_1)$ ratio of the ions of interest and contaminants, assuming an ion bunch width of 4~ns. For example, with a mass resolving power of $\approx 1\cdot 10^6$, contaminants with $m_1/(m_2-m_1)\approx  (1$ to $2)\cdot 10^5$ can be removed. This is sufficient for the scientific goals outlined in Sec.~\ref{sec:Motivation}. If a higher mass separation power is needed, the ion-bunch width can be increased. With a bunch width of $\approx 70$~ns a mass resolving power 1.5 to 3 times larger than the $m_1/(m_2-m_1)$ ratio is sufficient~\cite{WOLF2013123}, allowing removal of contaminants with $m_1/(m_2-m_1)=(3$ to $6)\cdot 10^5$ when again assuming a mass resolving power of $\approx 1\cdot 10^6$. Simulations show that by changing the potentials applied to the push and pull electrodes and the second cone of the Paul trap, the time focus can be matched to the position of the MR-ToF device for all tested initial time spreads between 3.7 and 150~ns. Across this range, the simulated mass resolving power in ideal conditions is above $10^7$. However, this improvement in BNG separation power comes at the expense of longer storage times.

\begin{figure}[t]
\centering
\includegraphics[width=\columnwidth]{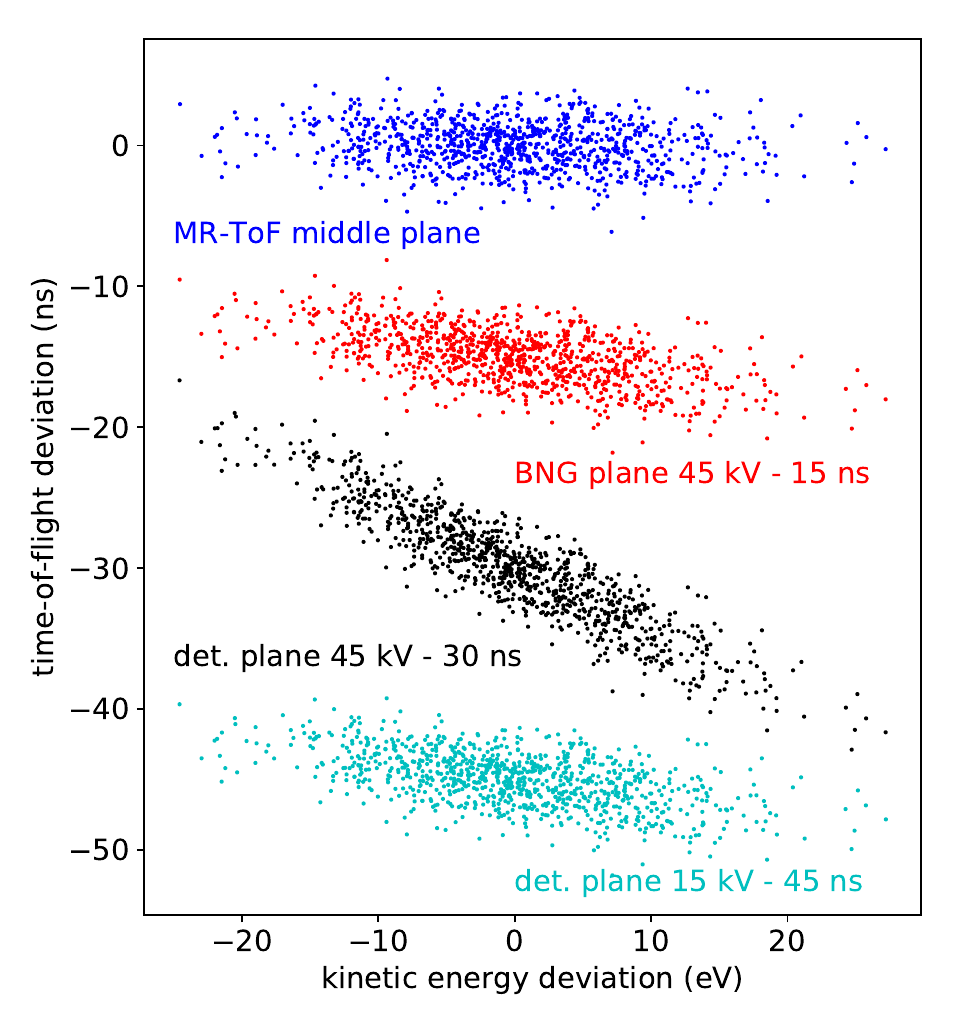}
\caption{Deviations in time-of-flight versus deviations in kinetic energy recorded at the MR-ToF middle plane, the BNG plane and the detector plane with the BNG biased to a floating voltage of either 15 or 45 kV. The ions are stored for 100 revolutions in the MR-ToF device prior to their extraction and their recording at the respective planes.}
\label{fig: ToFvsEforBNG}
\end{figure}

\subsubsection{Deflection electrodes}
In addition to mass-selective ion ejection and the use of a BNG, azimuthally segmented deflection electrodes can also be employed for the removal of contaminants. These electrodes enable in situ elimination of undesired ions while the remaining species continue to be stored within the MR-ToF device. To remove all contaminants within 10 revolutions, a voltage of $\pm 600$~V must be applied to two opposing segments of the deflector electrode. The time resolution of the deflector electrode is $\approx 480$ ns\footnote{The time resolution is determined by the time the ions are affected by the potential applied to the deflector electrode. A comparison of the resulting potential distribution with 0~V and $\pm 600$~V applied to the deflector electrode segments reveals a potential difference exceeding 0.05~V over a region of approximately 100~mm (considering radial distances up to 5~mm from the MR-ToF axis). At a beam energy of 30~keV, ions take $\approx 480$~ns to traverse this distance.} and thus comparable to that of the mass-selective ion ejection technique.
Activating the deflector electrodes allows for the removal of contaminants with significantly different masses than the ions of interest, thereby eliminating ambiguities in the assignment of revolution number. Furthermore, the ion density within the MR-ToF device can be reduced, mitigating space-charge effects during subsequent ion storage.

\subsubsection{Mass-selective retrapping}
The in the MR-ToF device mass-separated ions can also be guided back into the Paul-trap cooler-buncher for contaminant removal via mass-selective retrapping~\cite{doi:10.1007/s13361-017-1617-z}.
For example, the TITAN MR-ToF collaboration successfully performed an MR-ToF mass measurement of $^{60}$Ga despite receiving only 0.025 $^{60}$Ga ions per second out of a total beam rate of 10,000 ions per second~\cite{PhysRevC.104.065803}. To achieve this, they initially performed an MR-ToF mass separation with a resolving power of $m/(\Delta m)< 10^4$ sufficiently high to separate $^{60}$Ga from the contaminants. The ions were then selectively transferred back into the Paul trap to remove a large fraction of the contaminants. Since a significantly lower number of ions were then stored in the Paul trap, space charge effects in the Paul trap were reduced and the remaining ions with a smaller emittance could be injected back into the MR-ToF device. Due to the reduced number of ions injected into the MR-ToF device, also space-charge effects within the MR-ToF device were minimized and a high mass resolving power was achieved. 

In summary, the BNG will enable the removal of contaminants with the smallest time resolution $\tau_{\mathrm{res}}$ and thus the highest mass separation power. Prior to extracting the ions from the MR-ToF device toward the BNG, it might be beneficial to use the deflector electrodes and/or the mass-selective retrapping technique for an in-situ removal of distant contaminants. This would enhance the mass resolving power of the MR-ToF system before ion extraction from the MR-ToF device toward the BNG.

\subsection{\label{sec: Reduced Beam Energy} Performance For Ion Storage at 15 or 3 keV Beam Energy} We studied the performance characteristics of FRIB's MR-ToF device for 15 and 3~keV beam energies of the stored ions as well. In order to store ions at 15 keV beam energy, the transport beam energy needs to be reduced from 50~keV to $\approx$ 30 keV. The reason for this is that suitable high-voltage switches for the in-trap lift with fast time constants max out at 20 kV. The bias voltage of the Paul trap platform is hence also reduced to $\approx$ 30~kV. For ion storage at 3 keV beam energy in the MR-ToF device, the transport energy needs to be further reduced to $\approx$ 20~keV. The potentials applied to the MR-ToF mirror electrodes were scaled based on the reduction in beam energy. For 15~keV beam energy the simulated mass resolving power in ideal conditions drops down to $R_{\mathrm{inf}} = 8\cdot10^6$ and for 3~keV beam energy to $2\cdot10^6$. This mass resolving power is still high enough for the applications outlined in Sec.~\ref{sec:Motivation}. 

\begin{figure}[t]
\centering
\includegraphics[width=\columnwidth]{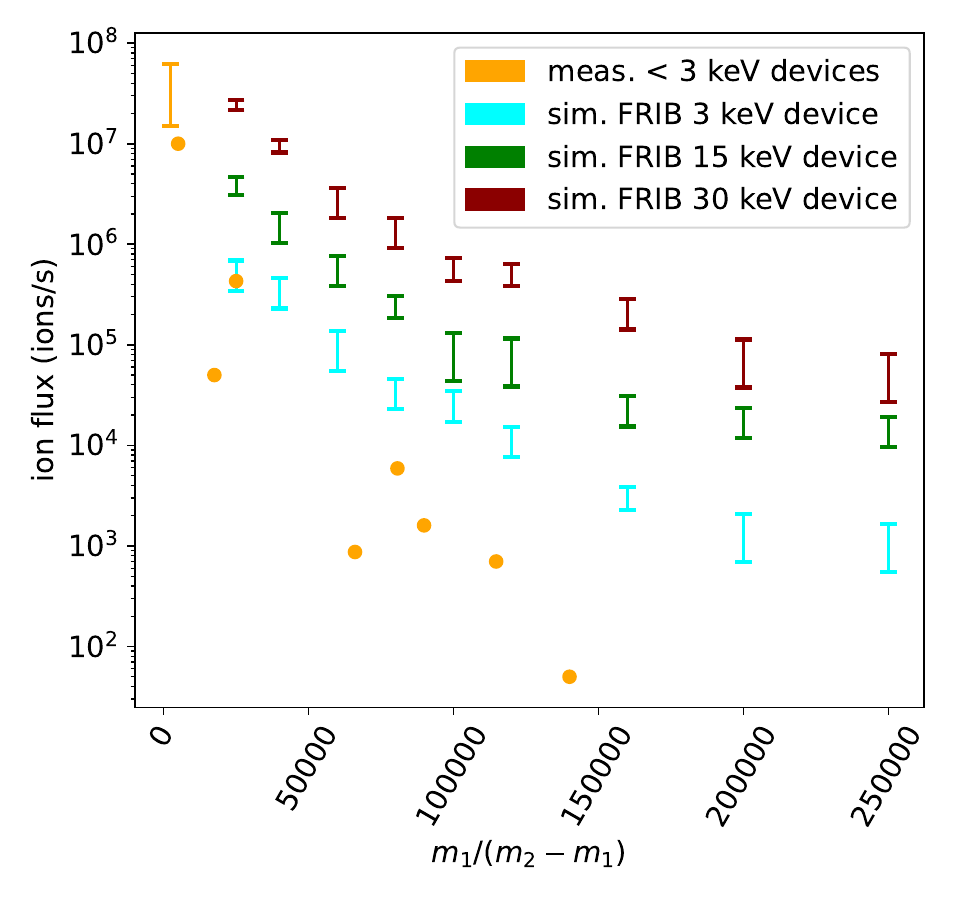}
\caption{Maximal ion flux versus $m_1/(m_2-m_1)$ for FRIB's MR-ToF MS for the three different simulation scenarios resulting in an ion storage energy of 3, 15 and 30 keV. The experimental ion flux values of existing $< 3$ keV low-energy MR-ToF MSs are shown in comparison, see Fig.~\ref{fig: IOn flux} for details. }
\label{fig: IOn flux_DiffBeamE}
\end{figure}

As discussed in Ref.~\cite{MAIER2023168545} the ion flux increases linearly with beam energy when using the same MR-ToF design and MR-ToF tuning mode (i.e., when scaling all potentials by the same factor). Due to the larger energy spread tolerance of a 30 keV MR-ToF device and the consequently reduced cross-order terms between energy and radial/angular deviations, combined with the large radial openings of all MR-ToF electrodes, it is, however, possible to detune the transverse focus of the ions while still maintaining a high mass resolving power and 100\% storage efficiency. In this way, the ions are further spread apart in the mirrors, which reduces space-charge effects and yields an additional increase in ion flux by a factor of $\approx 3$, as already discussed in Sec.~\ref{sec: Ion Flux}. Thus, the ion flux is a factor $\approx 30$ larger when storing ions at 30 instead of 3 keV in FRIB's MR-ToF MS.   Fig.~\ref{fig: IOn flux_DiffBeamE} shows the simulated ion flux as a function of $m_1/(m_2-m_1)$ for FRIB's MR-ToF MS for 3, 15, and 30 keV beam energy in comparison to the experimental ion-flux values of existing low-energy MR-ToF MSs storing ions at less than 3 keV beam energy.  Why the ion flux for FRIB's MR-ToF MS operated at 3 keV beam energy is also, on average, an order of magnitude higher than the ion flux of other low-energy MR-ToF MSs remains an open question. We attribute this to our optimization of the MR-ToF design and applied bias voltages in favor of high ion flux. For a beam energy of 3~keV, the ion flux remains above $10^4$ ions/s for $m_1/(m_2-m_1) <10^5$ which would already be largely beneficial for many measurements at FRIB. A gradual ramp-up of the stored energy in the MR-ToF device while gaining operational experience is hence a viable option.

\section{Conclusion and Outlook}
We have presented the motivation, conceptual design, and expected performance of a high-voltage MR-ToF mass spectrometer and separator (MR-ToF MS) under development at FRIB. This device, which is expected to store ions at 30 keV beam energy, is engineered to meet the growing demands of FRIB’s stopped and reaccelerated beam programs.
Extensive ion-optical simulations, starting from buffer-gas cooling and bunching in the Paul trap until ion injection, through storage in the MR-ToF device, and ending with extraction, have guided the design and voltage configuration of the system. 
Simulations indicate that mass separation powers exceeding $10^5$ can be achieved for an ion flux as high as $10^5$ ions/s, enabling the delivery of isobarically and isomerically purified ion beams at high ion intensities to subsequent experiments. This simulated ion flux is up to two orders of magnitude greater than that of state-of-the-art low-energy MR-ToF MS due to the combination of higher kinetic energy and optimized geometry. Under ideal conditions, the simulated mass resolving power exceeds $10^7$. Even under more realistic experimental conditions, such as including power supply instabilities of current state-of-the-art power supplies, a mass resolving power of $8\cdot 10^5$ is expected, which is sufficient to achieve the desired purification and mass measurement capabilities at FRIB.
Overall, the MR-ToF MS is poised to become a cornerstone of FRIB's stopped and reaccelerated beam facility, enabling high-precision mass spectrometry, fast beam diagnostics, and the delivery of isobarically and isomerically purified beams. 
Looking ahead, the next steps involve the construction and commissioning of the MR-ToF MS with offline ion sources, detailed experimental benchmarking, and eventual integration into the experimental beamlines. With its enhanced mass resolution and ion flux capabilities, the MR-ToF MS will unlock new experimental possibilities and help advance our understanding of the nuclear landscape.

\section*{\label{sec:Acknowledgments}Acknowledgments}
 This work was supported by the U.S. Department of Energy, Office of Science, Office of Nuclear Physics under Grant No. DE-SC0023633 (MSU) and DE-AC02-05CH11231. C.M.I. acknowledges support from the ASET Traineeship under the DOE award no. DE-SC0018362. This work was supported by the U.S. Army DEVCOM ARL Army Research Office (ARO) Energy Sciences Competency, (Electrochemistry or Advanced Energy Materials) Program award \# W911NF2510038. The views and conclusions contained in this document are those of the authors and should not be interpreted as representing the official policies, either expressed or implied, of the U.S. Army or the U.S. Government. This work was supported in part through computational resources and services provided by the Institute for Cyber-Enabled Research at Michigan State University.

\section*{\label{Appendix1}Appendix 1}
The potentials applied to the individual electrodes of the MR-ToF device are stated in Tab.~\ref{tab:MRToFSetting}. 
\begin{table}[h]
\centering
 \caption{\label{tab:MRToFSetting} Potentials applied to the MR-ToF mirror electrodes. The positions of the respective electrodes are shown in Fig.~\ref{fig: MR-ToF setup}.}
 \vspace{1mm}
 \begin{tabular}{cc}
   \hline
   \hline 
  respective electrode & potential (V)\\
 \hline
  1 (innermost electrode) &  -12\,908.4 \\
  2 &  -14\,348.2 \\
  3 & -1\,147.0 \\
  4 & 12\,771.5 \\
  5 &  21\,564.2 \\
  6 & 25\,329.3\\
  7 &  32\,384.3 \\
  8 &  36\,144.2 \\
  9 (outermost electrode) & 43\,340.5\\
  \hline
  \hline
 \end{tabular}
 \end{table}
\section*{\label{Appendix1}Appendix 2}
\begin{figure}[t]
\centering
\includegraphics[width=\columnwidth]{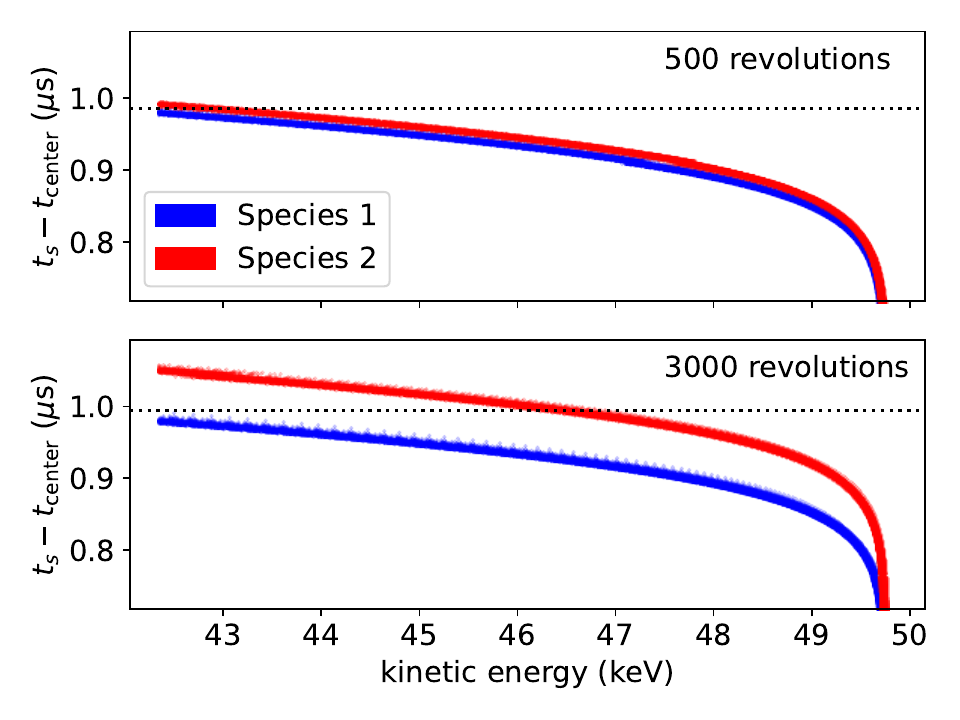}
\caption{Kinetic energy distribution for two simultaneously stored ion species with $A=132$~u and $m_1/(m_2-m_1) = 2\cdot 10^5$ 
for different storage times $t_s$ after which the in-trap lift is activated for ion extraction. Top: The ions are stored for 500 revolutions before the in-trap lift is activated ($t_\mathrm{center} \approx 4.6$~ms), resulting in $R\approx 8\cdot 10^5$. Bottom: The ions are stored for 3000 revolutions before the in-trap lift is activated ($t_\mathrm{center} \approx 27.6$~ms) resulting in $R\approx 4\cdot 10^6$. }
\label{fig: mass sel ion eject}
\end{figure}

In the following we provide more details on the simulations of the performance of the mass-selective ion ejection technique for FRIB's high-voltage MR-ToF MS. 
Fig.~\ref{fig: mass sel ion eject} shows the kinetic energy distribution of two simultaneously stored ion species with $A=132$~u and $m_1/(m_2-m_1) = 2\cdot 10^5$ for different storage times $t_s$ after which the in-trap lift is activated for ion extraction. Both ion species have a bunch width of 3.7~ns prior to the activation of the in-trap lift for ion ejection. The top panel depicts the case for 500 revolutions, for which the required storage time to extract the ions of interest when they are exactly in the center of the drift tube, $t_\mathrm{center}$, is $\approx$ 4.6~ms and a mass resolving power of $R \approx 8\cdot 10^5$ is reached. The bottom panel shows the case for 3000 revolutions for which $t_\mathrm{center} \approx 27.6$~ms and $R \approx 4\cdot 10^6$. If $t_s-t_\mathrm{center}$ is slightly larger than the dotted black line, only the heavier (species 2 with mass $m_2$) of the two simultaneously stored ion species can be extracted. To only extract the ions of interest, the $m_1/(m_2-m_1)$ ratio between the ions of interest and the contaminants must hence be at least a factor 4 smaller than the mass resolving power $R$ of the device for the given storage time. However, as can be seen in the top panel of Fig.~\ref{fig: mass sel ion eject}, the beam energy of the ions of interest is reduced by $\approx$ 6.9~keV in comparison to the case where the in-trap lift is switched when the ions of interest are within its center, so when $t_s-t_\mathrm{center}=0$. Also, the bunch width is increased to 220 ns compared to the 3.7 ns bunch width the ions would have when switching the in-trap lift when the ions are within its field-free region.  Even when the $m_1/(m_2-m_1)$ ratio between the ions of interest and the contaminating species is a factor of 20 smaller than $R$ (see bottom panel in Fig.~\ref{fig: mass sel ion eject}), the ions of interest are still disturbed by the switching of the in-trap lift. The beam energy of the ions of interest is reduced by 3.2~keV and the bunch width is increased to 30 ns. While the increased bunch width is not an issue for any of the subsequent experiments, the large energy differences result in additional beam tuning and optimization time. To leave the ions of interest fully undisturbed, the contaminants need to be $\approx 530$~ns separated from the ions of interest. This value is determined by the time the ions need at 30 keV beam energy to pass the 110 mm distance during which the potential curve for ion extraction (blue curve in Fig.~\ref{fig: MR-ToF setup}) climbs from 0 V to 18915~V within a potential difference of 0.05~V. A time resolution of $\tau_{\mathrm{res}}\approx530$~ns means that the $m_1/(m_2-m_1)$ ratio needs to be $\approx$ 130 times smaller than the mass resolving power $R$ at the given storage time for ion bunch widths of 4~ns (FWHM). While the mass-selective ion ejection technique does not require any further equipment, it either severely reduces the mass separation power of the MR-ToF system or requires additional optimization time for the transfer of the ion bunch to subsequent experiments due to the severe change in ion beam energy. 

\bibliography{biblio}

\begin{thebibliography}{10}
\expandafter\ifx\csname url\endcsname\relax
  \def\url#1{\texttt{#1}}\fi
\expandafter\ifx\csname urlprefix\endcsname\relax\def\urlprefix{URL }\fi
\expandafter\ifx\csname href\endcsname\relax
  \def\href#1#2{#2} \def\path#1{#1}\fi

\bibitem{VILLARI2023350}
A.~Villari, G.~Bollen, A.~Henriques, A.~Lapierre, S.~Nash, R.~Ringle, S.~Schwarz, C.~Sumithrarachchi, \href{https://www.sciencedirect.com/science/article/pii/S0168583X23002422}{{Gas stopping and reacceleration techniques at the Facility for Rare Isotope Beams (FRIB)}}, Nuclear Instruments and Methods in Physics Research Section B: Beam Interactions with Materials and Atoms 541 (2023) 350--354.
\newblock \href {https://doi.org/https://doi.org/10.1016/j.nimb.2023.05.037} {\path{doi:https://doi.org/10.1016/j.nimb.2023.05.037}}.
\newline\urlprefix\url{https://www.sciencedirect.com/science/article/pii/S0168583X23002422}

\bibitem{SUMITHRARACHCHI2020305}
C.~Sumithrarachchi, D.~Morrissey, S.~Schwarz, K.~Lund, G.~Bollen, R.~Ringle, G.~Savard, A.~Villari, \href{https://www.sciencedirect.com/science/article/pii/S0168583X19302617}{Beam thermalization in a large gas catcher}, Nuclear Instruments and Methods in Physics Research Section B: Beam Interactions with Materials and Atoms 463 (2020) 305--309.
\newblock \href {https://doi.org/https://doi.org/10.1016/j.nimb.2019.04.077} {\path{doi:https://doi.org/10.1016/j.nimb.2019.04.077}}.
\newline\urlprefix\url{https://www.sciencedirect.com/science/article/pii/S0168583X19302617}

\bibitem{LUND2020378}
K.~Lund, G.~Bollen, D.~Lawton, D.~Morrissey, J.~Ottarson, R.~Ringle, S.~Schwarz, C.~Sumithrarachchi, A.~Villari, J.~Yurkon, \href{https://www.sciencedirect.com/science/article/pii/S0168583X1930237X}{{Online tests of the Advanced Cryogenic Gas Stopper at NSCL}}, Nuclear Instruments and Methods in Physics Research Section B: Beam Interactions with Materials and Atoms 463 (2020) 378--381.
\newblock \href {https://doi.org/https://doi.org/10.1016/j.nimb.2019.04.053} {\path{doi:https://doi.org/10.1016/j.nimb.2019.04.053}}.
\newline\urlprefix\url{https://www.sciencedirect.com/science/article/pii/S0168583X1930237X}

\bibitem{PLASS2013}
W.~R. Plaß, T.~Dickel, C.~Scheidenberger, \href{https://www.sciencedirect.com/science/article/pii/S138738061300239X}{{Multiple-reflection time-of-flight mass spectrometry}}, International Journal of Mass Spectrometry 349-350 (2013) 134--144, 100 years of Mass Spectrometry.
\newblock \href {https://doi.org/https://doi.org/10.1016/j.ijms.2013.06.005} {\path{doi:https://doi.org/10.1016/j.ijms.2013.06.005}}.
\newline\urlprefix\url{https://www.sciencedirect.com/science/article/pii/S138738061300239X}

\bibitem{Schury2009}
P.~Schury, K.~Okada, S.~Shchepunov, T.~Sonoda, A.~Takamine, M.~Wada, H.~Wollnik, Y.~Yamazaki, \href{https://doi.org/10.1140/epja/i2009-10882-6}{Multi-reflection time-of-flight mass spectrograph for short-lived radioactive ions}, The European Physical Journal A 42~(3) (2009) 343.
\newblock \href {https://doi.org/10.1140/epja/i2009-10882-6} {\path{doi:10.1140/epja/i2009-10882-6}}.
\newline\urlprefix\url{https://doi.org/10.1140/epja/i2009-10882-6}

\bibitem{Wienholtz2013}
F.~Wienholtz, D.~Beck, K.~Blaum, C.~Borgmann, M.~Breitenfeldt, R.~B. Cakirli, S.~George, F.~Herfurth, J.~D. Holt, M.~Kowalska, S.~Kreim, D.~Lunney, V.~Manea, J.~Men{\'e}ndez, D.~Neidherr, M.~Rosenbusch, L.~Schweikhard, A.~Schwenk, J.~Simonis, J.~Stanja, R.~N. Wolf, K.~Zuber, \href{https://doi.org/10.1038/nature12226}{Masses of exotic calcium isotopes pin down nuclear forces}, Nature 498~(7454) (2013) 346--349.
\newblock \href {https://doi.org/10.1038/nature12226} {\path{doi:10.1038/nature12226}}.
\newline\urlprefix\url{https://doi.org/10.1038/nature12226}

\bibitem{REITER2021165823}
M.~Reiter, S.~A.~S. Andrés, J.~Bergmann, T.~Dickel, J.~Dilling, A.~Jacobs, A.~Kwiatkowski, W.~Plaß, C.~Scheidenberger, D.~Short, C.~Will, C.~Babcock, E.~Dunling, A.~Finlay, C.~Hornung, C.~Jesch, R.~Klawitter, B.~Kootte, D.~Lascar, E.~Leistenschneider, T.~Murböck, S.~Paul, M.~Yavor, \href{https://www.sciencedirect.com/science/article/pii/S0168900221008081}{{Commissioning and performance of TITAN’s Multiple-Reflection Time-of-Flight Mass-Spectrometer and isobar separator}}, Nuclear Instruments and Methods in Physics Research Section A: Accelerators, Spectrometers, Detectors and Associated Equipment 1018 (2021) 165823.
\newblock \href {https://doi.org/https://doi.org/10.1016/j.nima.2021.165823} {\path{doi:https://doi.org/10.1016/j.nima.2021.165823}}.
\newline\urlprefix\url{https://www.sciencedirect.com/science/article/pii/S0168900221008081}

\bibitem{CHAUVEAU2016211}
P.~Chauveau, P.~Delahaye, G.~{De France}, S.~{El Abir}, J.~Lory, Y.~Merrer, M.~Rosenbusch, L.~Schweikhard, R.~Wolf, \href{http://www.sciencedirect.com/science/article/pii/S0168583X16000732}{{PILGRIM, a Multi-Reflection Time-of-Flight Mass Spectrometer for Spiral2-S3 at GANIL}}, Nuclear Instruments and Methods in Physics Research Section B: Beam Interactions with Materials and Atoms 376 (2016) 211 -- 215, proceedings of the XVIIth International Conference on Electromagnetic Isotope Separators and Related Topics (EMIS2015), Grand Rapids, MI, U.S.A., 11-15 May 2015.
\newblock \href {https://doi.org/https://doi.org/10.1016/j.nimb.2016.01.025} {\path{doi:https://doi.org/10.1016/j.nimb.2016.01.025}}.
\newline\urlprefix\url{http://www.sciencedirect.com/science/article/pii/S0168583X16000732}

\bibitem{LIU2021164679}
B.~Liu, M.~Brodeur, D.~Burdette, J.~Kelly, T.~Kim, J.~Long, P.~O'Malley, \href{http://www.sciencedirect.com/science/article/pii/S0168900220310767}{The performance of the commissioned {Notre Dame} multi-reflection time-of-flight mass spectrometer}, Nuclear Instruments and Methods in Physics Research Section A: Accelerators, Spectrometers, Detectors and Associated Equipment 985 (2021) 164679.
\newblock \href {https://doi.org/https://doi.org/10.1016/j.nima.2020.164679} {\path{doi:https://doi.org/10.1016/j.nima.2020.164679}}.
\newline\urlprefix\url{http://www.sciencedirect.com/science/article/pii/S0168900220310767}

\bibitem{ROSENBUSCH2022167824}
M.~Rosenbusch, M.~Wada, S.~Chen, A.~Takamine, S.~Iimura, D.~Hou, W.~Xian, S.~Yan, P.~Schury, Y.~Hirayama, Y.~Ito, H.~Ishiyama, S.~Kimura, T.~Kojima, J.~Lee, J.~Liu, S.~Michimasa, H.~Miyatake, J.~Moon, M.~Mukai, S.~Nishimura, S.~Naimi, T.~Niwase, T.~Sonoda, Y.~Watanabe, H.~Wollnik, \href{https://www.sciencedirect.com/science/article/pii/S0168900222011160}{{The new MRTOF mass spectrograph following the ZeroDegree spectrometer at RIKEN’s RIBF facility}}, Nuclear Instruments and Methods in Physics Research Section A: Accelerators, Spectrometers, Detectors and Associated Equipment (2022) 167824\href {https://doi.org/https://doi.org/10.1016/j.nima.2022.167824} {\path{doi:https://doi.org/10.1016/j.nima.2022.167824}}.
\newline\urlprefix\url{https://www.sciencedirect.com/science/article/pii/S0168900222011160}

\bibitem{virtanen2025highresolutionmultireflectiontimeofflightmass}
V.~A. Virtanen, T.~Eronen, A.~Kankainen, O.~Beliuskina, Z.~Ge, R.~P. de~Groote, A.~Jokinen, M.~Mougeot, A.~de~Roubin, J.~Ruotsalainen, J.~Sarén, A.~Takkinen, \href{https://arxiv.org/abs/2508.10048}{{High-resolution multi-reflection time-of-flight mass spectrometer for exotic nuclei at IGISOL}} (2025).
\newblock \href {http://arxiv.org/abs/2508.10048} {\path{arXiv:2508.10048}}.
\newline\urlprefix\url{https://arxiv.org/abs/2508.10048}

\bibitem{DICKEL2015172}
T.~Dickel, W.~Pla{\ss}, A.~Becker, U.~Czok, H.~Geissel, E.~Haettner, C.~Jesch, W.~Kinsel, M.~Petrick, C.~Scheidenberger, A.~Simon, M.~Yavor, \href{http://www.sciencedirect.com/science/article/pii/S0168900214015629}{A high-performance multiple-reflection time-of-flight mass spectrometer and isobar separator for the research with exotic nuclei}, Nuclear Instruments and Methods in Physics Research Section A: Accelerators, Spectrometers, Detectors and Associated Equipment 777 (2015) 172 -- 188.
\newblock \href {https://doi.org/https://doi.org/10.1016/j.nima.2014.12.094} {\path{doi:https://doi.org/10.1016/j.nima.2014.12.094}}.
\newline\urlprefix\url{http://www.sciencedirect.com/science/article/pii/S0168900214015629}

\bibitem{PhysRev.49.388}
N.~E. Bradbury, R.~A. Nielsen, \href{https://link.aps.org/doi/10.1103/PhysRev.49.388}{Absolute values of the electron mobility in hydrogen}, Phys. Rev. 49 (1936) 388--393.
\newblock \href {https://doi.org/10.1103/PhysRev.49.388} {\path{doi:10.1103/PhysRev.49.388}}.
\newline\urlprefix\url{https://link.aps.org/doi/10.1103/PhysRev.49.388}

\bibitem{PLA20084560}
W.~R. Plaß, T.~Dickel, U.~Czok, H.~Geissel, M.~Petrick, K.~Reinheimer, C.~Scheidenberger, M.~I.Yavor, \href{https://www.sciencedirect.com/science/article/pii/S0168583X08007763}{Isobar separation by time-of-flight mass spectrometry for low-energy radioactive ion beam facilities}, Nuclear Instruments and Methods in Physics Research Section B: Beam Interactions with Materials and Atoms 266~(19) (2008) 4560--4564, proceedings of the XVth International Conference on Electromagnetic Isotope Separators and Techniques Related to their Applications.
\newblock \href {https://doi.org/https://doi.org/10.1016/j.nimb.2008.05.079} {\path{doi:https://doi.org/10.1016/j.nimb.2008.05.079}}.
\newline\urlprefix\url{https://www.sciencedirect.com/science/article/pii/S0168583X08007763}

\bibitem{WOLF201282}
R.~Wolf, D.~Beck, K.~Blaum, C.~Böhm, C.~Borgmann, M.~Breitenfeldt, F.~Herfurth, A.~Herlert, M.~Kowalska, S.~Kreim, D.~Lunney, S.~Naimi, D.~Neidherr, M.~Rosenbusch, L.~Schweikhard, J.~Stanja, F.~Wienholtz, K.~Zuber, \href{https://www.sciencedirect.com/science/article/pii/S016890021200575X}{On-line separation of short-lived nuclei by a multi-reflection time-of-flight device}, Nuclear Instruments and Methods in Physics Research Section A: Accelerators, Spectrometers, Detectors and Associated Equipment 686 (2012) 82--90.
\newblock \href {https://doi.org/https://doi.org/10.1016/j.nima.2012.05.067} {\path{doi:https://doi.org/10.1016/j.nima.2012.05.067}}.
\newline\urlprefix\url{https://www.sciencedirect.com/science/article/pii/S016890021200575X}

\bibitem{MAIER2023168545}
F.~M. Maier, F.~Buchinger, L.~Croquette, P.~Fischer, H.~Heylen, F.~Hummer, C.~Kanitz, A.~A. Kwiatkowski, V.~Lagaki, S.~Lechner, E.~Leistenschneider, G.~Neyens, P.~Plattner, A.~Roitman, M.~Rosenbusch, L.~Schweikhard, S.~Sels, M.~Vilen, F.~Wienholtz, S.~Malbrunot-Ettenauer, \href{https://www.sciencedirect.com/science/article/pii/S0168900223005351}{{Increased beam energy as a pathway towards a highly selective and high-flux MR-ToF mass separator}}, Nuclear Instruments and Methods in Physics Research Section A: Accelerators, Spectrometers, Detectors and Associated Equipment 1056 (2023) 168545.
\newblock \href {https://doi.org/https://doi.org/10.1016/j.nima.2023.168545} {\path{doi:https://doi.org/10.1016/j.nima.2023.168545}}.
\newline\urlprefix\url{https://www.sciencedirect.com/science/article/pii/S0168900223005351}

\bibitem{YAVOR20181}
M.~I. Yavor, T.~V. Pomozov, S.~N. Kirillov, Y.~I. Khasin, A.~N. Verenchikov, \href{https://www.sciencedirect.com/science/article/pii/S1387380617304840}{High performance gridless ion mirrors for multi-reflection time-of-flight and electrostatic trap mass analyzers}, International Journal of Mass Spectrometry 426 (2018) 1--11.
\newblock \href {https://doi.org/https://doi.org/10.1016/j.ijms.2018.01.009} {\path{doi:https://doi.org/10.1016/j.ijms.2018.01.009}}.
\newline\urlprefix\url{https://www.sciencedirect.com/science/article/pii/S1387380617304840}

\bibitem{YAMAGUCHI2021103882}
T.~Yamaguchi, H.~Koura, Y.~Litvinov, M.~Wang, \href{https://www.sciencedirect.com/science/article/pii/S0146641021000363}{Masses of exotic nuclei}, Progress in Particle and Nuclear Physics 120 (2021) 103882.
\newblock \href {https://doi.org/https://doi.org/10.1016/j.ppnp.2021.103882} {\path{doi:https://doi.org/10.1016/j.ppnp.2021.103882}}.
\newline\urlprefix\url{https://www.sciencedirect.com/science/article/pii/S0146641021000363}

\bibitem{Blaum2024}
K.~Blaum, M.~J.~G. Borge, \href{https://doi.org/10.1140/epja/s10050-024-01274-x}{Precision nuclear physics experiments and theory}, The European Physical Journal A 60~(4) (2024) 94.
\newblock \href {https://doi.org/10.1140/epja/s10050-024-01274-x} {\path{doi:10.1140/epja/s10050-024-01274-x}}.
\newline\urlprefix\url{https://doi.org/10.1140/epja/s10050-024-01274-x}

\bibitem{Clark2023}
J.~Clark, G.~Savard, M.~Mumpower, A.~Kankainen, \href{https://doi.org/10.1140/epja/s10050-023-01037-0}{Precise mass measurements of radioactive nuclides for astrophysics}, The European Physical Journal A 59~(9) (2023) 204.
\newblock \href {https://doi.org/10.1140/epja/s10050-023-01037-0} {\path{doi:10.1140/epja/s10050-023-01037-0}}.
\newline\urlprefix\url{https://doi.org/10.1140/epja/s10050-023-01037-0}

\bibitem{mumpower2016impact}
M.~Mumpower, R.~Surman, G.~McLaughlin, A.~Aprahamian, \href{https://www.sciencedirect.com/science/article/pii/S0146641015000897}{{The impact of individual nuclear properties on r-process nucleosynthesis}}, Progress in Particle and Nuclear Physics 86 (2016) 86--126.
\newline\urlprefix\url{https://www.sciencedirect.com/science/article/pii/S0146641015000897}

\bibitem{RINGLE201387}
R.~Ringle, S.~Schwarz, G.~Bollen, \href{https://www.sciencedirect.com/science/article/pii/S1387380613001279}{{Penning trap mass spectrometry of rare isotopes produced via projectile fragmentation at the LEBIT facility}}, International Journal of Mass Spectrometry 349-350 (2013) 87--93, 100 years of Mass Spectrometry.
\newblock \href {https://doi.org/https://doi.org/10.1016/j.ijms.2013.04.001} {\path{doi:https://doi.org/10.1016/j.ijms.2013.04.001}}.
\newline\urlprefix\url{https://www.sciencedirect.com/science/article/pii/S1387380613001279}

\bibitem{MATOS2012171}
M.~Matoš, A.~Estradé, H.~Schatz, D.~Bazin, M.~Famiano, A.~Gade, S.~George, W.~Lynch, Z.~Meisel, M.~Portillo, A.~Rogers, D.~Shapira, A.~Stolz, M.~Wallace, J.~Yurkon, \href{https://www.sciencedirect.com/science/article/pii/S0168900212010078}{Time-of-flight mass measurements of exotic nuclei}, Nuclear Instruments and Methods in Physics Research Section A: Accelerators, Spectrometers, Detectors and Associated Equipment 696 (2012) 171--179.
\newblock \href {https://doi.org/https://doi.org/10.1016/j.nima.2012.08.104} {\path{doi:https://doi.org/10.1016/j.nima.2012.08.104}}.
\newline\urlprefix\url{https://www.sciencedirect.com/science/article/pii/S0168900212010078}

\bibitem{MEISEL2013145}
Z.~Meisel, S.~George, \href{https://www.sciencedirect.com/science/article/pii/S1387380613001139}{Time-of-flight mass spectrometry of very exotic systems}, International Journal of Mass Spectrometry 349-350 (2013) 145--150, 100 years of Mass Spectrometry.
\newblock \href {https://doi.org/https://doi.org/10.1016/j.ijms.2013.03.022} {\path{doi:https://doi.org/10.1016/j.ijms.2013.03.022}}.
\newline\urlprefix\url{https://www.sciencedirect.com/science/article/pii/S1387380613001139}

\bibitem{cannarozzo2025isomericyieldratiosmass}
S.~Cannarozzo, S.~Pomp, A.~Kankainen, I.~Moore, M.~Stryjczyk, A.~Al-Adili, A.~Solders, V.~Virtanen, T.~Eronen, Z.~Gao, Z.~Ge, A.~Jaries, M.~Lantz, M.~Mougeot, H.~Penttilä, A.~Raggio, J.~Ruotsalainen, \href{https://arxiv.org/abs/2504.11274}{{Isomeric yield ratios and mass spectrometry of Y and Nb isotopes in the neutron-rich N=60 region: the unusual case of $^{98}$Y}} (2025).
\newblock \href {http://arxiv.org/abs/2504.11274} {\path{arXiv:2504.11274}}.
\newline\urlprefix\url{https://arxiv.org/abs/2504.11274}

\bibitem{Dickel2024}
T.~Dickel, A.~Mollaebrahimi, \href{https://doi.org/10.1140/epjs/s11734-024-01156-9}{Unveiling nuclear isomers through multiple-reflection time-of-flight mass spectrometry}, The European Physical Journal Special Topics 233~(5) (2024) 1181--1190.
\newblock \href {https://doi.org/10.1140/epjs/s11734-024-01156-9} {\path{doi:10.1140/epjs/s11734-024-01156-9}}.
\newline\urlprefix\url{https://doi.org/10.1140/epjs/s11734-024-01156-9}

\bibitem{PhysRevLett.124.092502}
V.~Manea, J.~Karthein, D.~Atanasov, M.~Bender, K.~Blaum, T.~E. Cocolios, S.~Eliseev, A.~Herlert, J.~D. Holt, W.~J. Huang, Y.~A. Litvinov, D.~Lunney, J.~Men\'endez, M.~Mougeot, D.~Neidherr, L.~Schweikhard, A.~Schwenk, J.~Simonis, A.~Welker, F.~Wienholtz, K.~Zuber, \href{https://link.aps.org/doi/10.1103/PhysRevLett.124.092502}{First glimpse of the $n=82$ shell closure below $z=50$ from masses of neutron-rich cadmium isotopes and isomers}, Phys. Rev. Lett. 124 (2020) 092502.
\newblock \href {https://doi.org/10.1103/PhysRevLett.124.092502} {\path{doi:10.1103/PhysRevLett.124.092502}}.
\newline\urlprefix\url{https://link.aps.org/doi/10.1103/PhysRevLett.124.092502}

\bibitem{Mougeot2021}
M.~Mougeot, D.~Atanasov, J.~Karthein, R.~N. Wolf, P.~Ascher, K.~Blaum, K.~Chrysalidis, G.~Hagen, J.~D. Holt, W.~J. Huang, G.~R. Jansen, I.~Kulikov, Y.~A. Litvinov, D.~Lunney, V.~Manea, T.~Miyagi, T.~Papenbrock, L.~Schweikhard, A.~Schwenk, T.~Steinsberger, S.~R. Stroberg, Z.~H. Sun, A.~Welker, F.~Wienholtz, S.~G. Wilkins, K.~Zuber, \href{https://doi.org/10.1038/s41567-021-01326-9}{Mass measurements of 99--101in challenge ab initio nuclear theory of the nuclide 100sn}, Nature Physics 17~(10) (2021) 1099--1103.
\newblock \href {https://doi.org/10.1038/s41567-021-01326-9} {\path{doi:10.1038/s41567-021-01326-9}}.
\newline\urlprefix\url{https://doi.org/10.1038/s41567-021-01326-9}

\bibitem{PhysRevLett.120.062503}
E.~Leistenschneider, M.~P. Reiter, S.~Ayet San~Andr\'es, B.~Kootte, J.~D. Holt, P.~Navr\'atil, C.~Babcock, C.~Barbieri, B.~R. Barquest, J.~Bergmann, J.~Bollig, T.~Brunner, E.~Dunling, A.~Finlay, H.~Geissel, L.~Graham, F.~Greiner, H.~Hergert, C.~Hornung, C.~Jesch, R.~Klawitter, Y.~Lan, D.~Lascar, K.~G. Leach, W.~Lippert, J.~E. McKay, S.~F. Paul, A.~Schwenk, D.~Short, J.~Simonis, V.~Som\`a, R.~Steinbr\"ugge, S.~R. Stroberg, R.~Thompson, M.~E. Wieser, C.~Will, M.~Yavor, C.~Andreoiu, T.~Dickel, I.~Dillmann, G.~Gwinner, W.~R. Pla\ss{}, C.~Scheidenberger, A.~A. Kwiatkowski, J.~Dilling, \href{https://link.aps.org/doi/10.1103/PhysRevLett.120.062503}{Dawning of the $n=32$ shell closure seen through precision mass measurements of neutron-rich titanium isotopes}, Phys. Rev. Lett. 120 (2018) 062503.
\newblock \href {https://doi.org/10.1103/PhysRevLett.120.062503} {\path{doi:10.1103/PhysRevLett.120.062503}}.
\newline\urlprefix\url{https://link.aps.org/doi/10.1103/PhysRevLett.120.062503}

\bibitem{ToF1}
G.~Bollen, R.~B. Moore, G.~Savard, H.~Stolzenberg, \href{https://doi.org/10.1063/1.346185}{{The accuracy of heavy‐ion mass measurements using time of flight‐ion cyclotron resonance in a Penning trap}}, Journal of Applied Physics 68~(9) (1990) 4355--4374.
\newblock \href {http://arxiv.org/abs/https://doi.org/10.1063/1.346185} {\path{arXiv:https://doi.org/10.1063/1.346185}}, \href {https://doi.org/10.1063/1.346185} {\path{doi:10.1063/1.346185}}.
\newline\urlprefix\url{https://doi.org/10.1063/1.346185}

\bibitem{BECKER199053}
S.~Becker, G.~Bollen, F.~Kern, H.-J. Kluge, R.~Moore, G.~Savard, L.~Schweikhard, H.~Stolzenberg, \href{https://www.sciencedirect.com/science/article/pii/016811769085021S}{{Mass measurements of very high accuracy by time-of-flight ion cyclotron resonance of ions injected into a Penning trap}}, International Journal of Mass Spectrometry and Ion Processes 99~(1) (1990) 53--77.
\newblock \href {https://doi.org/https://doi.org/10.1016/0168-1176(90)85021-S} {\path{doi:https://doi.org/10.1016/0168-1176(90)85021-S}}.
\newline\urlprefix\url{https://www.sciencedirect.com/science/article/pii/016811769085021S}

\bibitem{KONIG199595}
M.~König, G.~Bollen, H.-J. Kluge, T.~Otto, J.~Szerypo, \href{https://www.sciencedirect.com/science/article/pii/016811769504146C}{Quadrupole excitation of stored ion motion at the true cyclotron frequency}, International Journal of Mass Spectrometry and Ion Processes 142~(1) (1995) 95--116.
\newblock \href {https://doi.org/https://doi.org/10.1016/0168-1176(95)04146-C} {\path{doi:https://doi.org/10.1016/0168-1176(95)04146-C}}.
\newline\urlprefix\url{https://www.sciencedirect.com/science/article/pii/016811769504146C}

\bibitem{PhysRevLett.110.082501}
S.~Eliseev, K.~Blaum, M.~Block, C.~Droese, M.~Goncharov, E.~Minaya~Ramirez, D.~A. Nesterenko, Y.~N. Novikov, L.~Schweikhard, \href{https://link.aps.org/doi/10.1103/PhysRevLett.110.082501}{Phase-imaging ion-cyclotron-resonance measurements for short-lived nuclides}, Phys. Rev. Lett. 110 (2013) 082501.
\newblock \href {https://doi.org/10.1103/PhysRevLett.110.082501} {\path{doi:10.1103/PhysRevLett.110.082501}}.
\newline\urlprefix\url{https://link.aps.org/doi/10.1103/PhysRevLett.110.082501}

\bibitem{Eliseev2014}
S.~Eliseev, K.~Blaum, M.~Block, A.~D{\"o}rr, C.~Droese, T.~Eronen, M.~Goncharov, M.~H{\"o}cker, J.~Ketter, E.~M. Ramirez, D.~A. Nesterenko, Y.~N. Novikov, L.~Schweikhard, \href{https://doi.org/10.1007/s00340-013-5621-0}{A phase-imaging technique for cyclotron-frequency measurements}, Applied Physics B 114~(1) (2014) 107--128.
\newblock \href {https://doi.org/10.1007/s00340-013-5621-0} {\path{doi:10.1007/s00340-013-5621-0}}.
\newline\urlprefix\url{https://doi.org/10.1007/s00340-013-5621-0}

\bibitem{PhysRevC.103.034319}
I.~Mardor, S.~A.~S. Andr\'es, T.~Dickel, D.~Amanbayev, S.~Beck, J.~Bergmann, H.~Geissel, L.~Gr\"of, E.~Haettner, C.~Hornung, N.~Kalantar-Nayestanaki, G.~Kripko-Koncz, I.~Miskun, A.~Mollaebrahimi, W.~R. Pla\ss{}, C.~Scheidenberger, H.~Weick, S.~Bagchi, D.~L. Balabanski, A.~A. Bezbakh, Z.~Brencic, O.~Charviakova, V.~Chudoba, P.~Constantin, M.~Dehghan, A.~S. Fomichev, L.~V. Grigorenko, O.~Hall, M.~N. Harakeh, J.-P. Hucka, A.~Kankainen, O.~Kiselev, R.~Kn\"obel, D.~A. Kostyleva, S.~A. Krupko, N.~Kurkova, N.~Kuzminchuk, I.~Mukha, I.~A. Muzalevskii, D.~Nichita, C.~Nociforo, Z.~Patyk, M.~Pf\"utzner, S.~Pietri, S.~Purushothaman, M.~P. Reiter, H.~Roesch, F.~Schirru, P.~G. Sharov, A.~Sp\ifmmode~\u{a}\else \u{a}\fi{}taru, G.~Stanic, A.~State, Y.~K. Tanaka, M.~Vencelj, M.~I. Yavor, J.~Zhao, \href{https://link.aps.org/doi/10.1103/PhysRevC.103.034319}{{Mass measurements of As, Se, and Br nuclei, and their implication on the proton-neutron interaction strength toward the $N=Z$ line}}, Phys. Rev. C 103 (2021) 034319.
\newblock \href {https://doi.org/10.1103/PhysRevC.103.034319} {\path{doi:10.1103/PhysRevC.103.034319}}.
\newline\urlprefix\url{https://link.aps.org/doi/10.1103/PhysRevC.103.034319}

\bibitem{Bradcom}
{Personal Communications with Brad Sherrill, Associate Director for Fragment Separation,} (2025).

\bibitem{AME2020}
M.~Wang, W.~Huang, F.~Kondev, G.~Audi, S.~Naimi, \href{https://dx.doi.org/10.1088/1674-1137/abddaf}{{The AME 2020 atomic mass evaluation (II). Tables, graphs and references*}}, Chinese Physics C 45~(3) (2021) 030003.
\newblock \href {https://doi.org/10.1088/1674-1137/abddaf} {\path{doi:10.1088/1674-1137/abddaf}}.
\newline\urlprefix\url{https://dx.doi.org/10.1088/1674-1137/abddaf}

\bibitem{annurev:/content/journals/10.1146/annurev.astro.46.060407.145207}
C.~Sneden, J.~J. Cowan, R.~Gallino, \href{https://www.annualreviews.org/content/journals/10.1146/annurev.astro.46.060407.145207}{Neutron-capture elements in the early galaxy}, Annual Review of Astronomy and Astrophysics 46~(Volume 46, 2008) (2008) 241--288.
\newblock \href {https://doi.org/https://doi.org/10.1146/annurev.astro.46.060407.145207} {\path{doi:https://doi.org/10.1146/annurev.astro.46.060407.145207}}.
\newline\urlprefix\url{https://www.annualreviews.org/content/journals/10.1146/annurev.astro.46.060407.145207}

\bibitem{PhysRevLett.132.152501}
S.~E. Campbell, G.~Bollen, B.~A. Brown, A.~Dockery, C.~M. Ireland, K.~Minamisono, D.~Puentes, B.~J. Rickey, R.~Ringle, I.~T. Yandow, K.~Fossez, A.~Ortiz-Cortes, S.~Schwarz, C.~S. Sumithrarachchi, A.~C.~C. Villari, \href{https://link.aps.org/doi/10.1103/PhysRevLett.132.152501}{Precision mass measurement of the proton dripline halo candidate $^{22}\mathrm{Al}$}, Phys. Rev. Lett. 132 (2024) 152501.
\newblock \href {https://doi.org/10.1103/PhysRevLett.132.152501} {\path{doi:10.1103/PhysRevLett.132.152501}}.
\newline\urlprefix\url{https://link.aps.org/doi/10.1103/PhysRevLett.132.152501}

\bibitem{Wolf2012OnlineSO}
R.~N. Wolf, D.~Beck, K.~Blaum, C.~B{\"o}hm, C.~Borgmann, M.~Breitenfeldt, F.~Herfurth, A.~Herlert, M.~Kowalska, S.~Kreim, D.~Lunney, S.~Naimi, D.~Neidherr, M.~Rosenbusch, L.~Schweikhard, J.~Stanja, F.~Wienholtz, K.~Zuber, \href{https://www.sciencedirect.com/science/article/pii/S016890021200575X}{On-line separation of short-lived nuclei by a multi-reflection time-of-flight device}, Nuclear Instruments and Methods in Physics Research Section A 686 (2012) 82--90.
\newblock \href {https://doi.org/https://doi.org/10.1016/j.nima.2012.05.067} {\path{doi:https://doi.org/10.1016/j.nima.2012.05.067}}.
\newline\urlprefix\url{https://www.sciencedirect.com/science/article/pii/S016890021200575X}

\bibitem{WOLF2013123}
R.~Wolf, F.~Wienholtz, D.~Atanasov, D.~Beck, K.~Blaum, C.~Borgmann, F.~Herfurth, M.~Kowalska, S.~Kreim, Y.~A. Litvinov, D.~Lunney, V.~Manea, D.~Neidherr, M.~Rosenbusch, L.~Schweikhard, J.~Stanja, K.~Zuber, \href{http://www.sciencedirect.com/science/article/pii/S1387380613001115}{{ISOLTRAP's multi-reflection time-of-flight mass separator/spectrometer}}, International Journal of Mass Spectrometry 349-350 (2013) 123 -- 133, 100 years of Mass Spectrometry.
\newblock \href {https://doi.org/https://doi.org/10.1016/j.ijms.2013.03.020} {\path{doi:https://doi.org/10.1016/j.ijms.2013.03.020}}.
\newline\urlprefix\url{http://www.sciencedirect.com/science/article/pii/S1387380613001115}

\bibitem{PORTILLO2023151}
M.~Portillo, B.~Sherrill, Y.~Choi, M.~Cortesi, K.~Fukushima, M.~Hausmann, E.~Kwan, S.~Lidia, P.~Ostroumov, R.~Ringle, M.~Smith, M.~Steiner, O.~Tarasov, A.~Villari, T.~Zhang, \href{https://www.sciencedirect.com/science/article/pii/S0168583X23001556}{{Commissioning of the Advanced Rare Isotope Separator ARIS at FRIB}}, Nuclear Instruments and Methods in Physics Research Section B: Beam Interactions with Materials and Atoms 540 (2023) 151--157.
\newblock \href {https://doi.org/https://doi.org/10.1016/j.nimb.2023.04.025} {\path{doi:https://doi.org/10.1016/j.nimb.2023.04.025}}.
\newline\urlprefix\url{https://www.sciencedirect.com/science/article/pii/S0168583X23001556}

\bibitem{SUMITHRARACHCHI2023301}
C.~Sumithrarachchi, Y.~Liu, S.~Rogers, S.~Schwarz, G.~Bollen, N.~Gamage, A.~Henriques, A.~Lapierre, R.~Ringle, I.~Yandow, A.~Villari, K.~Domnanich, S.~Satija, G.~Severin, M.~Au, J.~Ballof, Y.~V. Garcia, M.~Owen, E.~Reis, S.~Rothe, S.~Stegemann, \href{https://www.sciencedirect.com/science/article/pii/S0168583X23002665}{{The new Batch Mode Ion Source for stand-alone operation at the Facility for Rare Isotope Beams (FRIB)}}, Nuclear Instruments and Methods in Physics Research Section B: Beam Interactions with Materials and Atoms 541 (2023) 301--304.
\newblock \href {https://doi.org/https://doi.org/10.1016/j.nimb.2023.05.061} {\path{doi:https://doi.org/10.1016/j.nimb.2023.05.061}}.
\newline\urlprefix\url{https://www.sciencedirect.com/science/article/pii/S0168583X23002665}

\bibitem{Abel_2019}
E.~P. Abel, M.~Avilov, V.~Ayres, E.~Birnbaum, G.~Bollen, G.~Bonito, T.~Bredeweg, H.~Clause, A.~Couture, J.~DeVore, M.~Dietrich, P.~Ellison, J.~Engle, R.~Ferrieri, J.~Fitzsimmons, M.~Friedman, D.~Georgobiani, S.~Graves, J.~Greene, S.~Lapi, C.~S. Loveless, T.~Mastren, C.~Martinez-Gomez, S.~McGuinness, W.~Mittig, D.~Morrissey, G.~Peaslee, F.~Pellemoine, J.~D. Robertson, N.~Scielzo, M.~Scott, G.~Severin, D.~Shaughnessy, J.~Shusterman, J.~Singh, M.~Stoyer, L.~Sutherlin, A.~Visser, J.~Wilkinson, \href{https://dx.doi.org/10.1088/1361-6471/ab26cc}{{Isotope harvesting at FRIB: additional opportunities for scientific discovery}}, Journal of Physics G: Nuclear and Particle Physics 46~(10) (2019) 100501.
\newblock \href {https://doi.org/10.1088/1361-6471/ab26cc} {\path{doi:10.1088/1361-6471/ab26cc}}.
\newline\urlprefix\url{https://dx.doi.org/10.1088/1361-6471/ab26cc}

\bibitem{Villari:IPAC2016-TUPMR024}
A.~Villari, et~al., \href{http://jacow.org/ipac2016/papers/tupmr024.pdf}{{C}ommissioning and {F}irst {A}ccelerated {B}eams in the {R}eaccelerator ({R}ea3) of the {N}ational {S}uperconducting {C}yclotron {L}aboratory, {MSU}}, in: Proc. of International Particle Accelerator Conference (IPAC'16), Busan, Korea, May 8-13, 2016, no.~7 in International Particle Accelerator Conference, JACoW, Geneva, Switzerland, 2016, pp. 1287--1290, doi:10.18429/JACoW-IPAC2016-TUPMR024.
\newblock \href {https://doi.org/doi:10.18429/JACoW-IPAC2016-TUPMR024} {\path{doi:doi:10.18429/JACoW-IPAC2016-TUPMR024}}.
\newline\urlprefix\url{http://jacow.org/ipac2016/papers/tupmr024.pdf}

\bibitem{SIMON201316}
A.~Simon, S.~Quinn, A.~Spyrou, A.~Battaglia, I.~Beskin, A.~Best, B.~Bucher, M.~Couder, P.~DeYoung, X.~Fang, J.~Görres, A.~Kontos, Q.~Li, S.~Liddick, A.~Long, S.~Lyons, K.~Padmanabhan, J.~Peace, A.~Roberts, D.~Robertson, K.~Smith, M.~Smith, E.~Stech, B.~Stefanek, W.~Tan, X.~Tang, M.~Wiescher, \href{https://www.sciencedirect.com/science/article/pii/S0168900212013824}{{SuN: Summing NaI(Tl) gamma-ray detector for capture reaction measurements}}, Nuclear Instruments and Methods in Physics Research Section A: Accelerators, Spectrometers, Detectors and Associated Equipment 703 (2013) 16--21.
\newblock \href {https://doi.org/https://doi.org/10.1016/j.nima.2012.11.045} {\path{doi:https://doi.org/10.1016/j.nima.2012.11.045}}.
\newline\urlprefix\url{https://www.sciencedirect.com/science/article/pii/S0168900212013824}

\bibitem{RONNING2026170930}
E.~Ronning, S.~Uthayakumaar, A.~Spyrou, A.~Richard, S.~Liddick, A.~Tsantiri, R.~Ringle, H.~Arora, H.~Berg, J.~Berkman, D.~Bleuel, K.~Bosmpotinis, S.~Campbell, X.~Chen, B.~Crider, R.~Coleman, P.~DeYoung, A.~Doetsch, H.~Erington, T.~Gaballah, N.~Gamage, E.~Good, B.~Greaves, A.~Hartley, J.~Huffman, C.~Ireland, C.~Izzo, R.~Jain, J.~Larsson, R.~Lubna, F.~Maier, M.~Mogannam, M.~Mumpower, G.~Owens-Fryar, T.~Ogunbeku, D.~Scriven, M.~Smith, C.~Sumithrarachchi, A.~Sweet, K.~Taft, M.~Wiedeking, \href{https://www.sciencedirect.com/science/article/pii/S0168900225007326}{{The upgraded summing NaI(Tl) (SuN++) absorption spectrometer}}, Nuclear Instruments and Methods in Physics Research Section A: Accelerators, Spectrometers, Detectors and Associated Equipment 1082 (2026) 170930.
\newblock \href {https://doi.org/https://doi.org/10.1016/j.nima.2025.170930} {\path{doi:https://doi.org/10.1016/j.nima.2025.170930}}.
\newline\urlprefix\url{https://www.sciencedirect.com/science/article/pii/S0168900225007326}

\bibitem{Ronning2025PLB}
E.~Ronning, {Total Absorption Spectroscopy of Two Isomers in 70Cu Influencing Nucleosynthesis Signatures}, Submitted to Phys. Let. B (2025).

\bibitem{RevModPhys.83.1467}
K.~Heyde, J.~L. Wood, \href{https://link.aps.org/doi/10.1103/RevModPhys.83.1467}{Shape coexistence in atomic nuclei}, Rev. Mod. Phys. 83 (2011) 1467--1521.
\newblock \href {https://doi.org/10.1103/RevModPhys.83.1467} {\path{doi:10.1103/RevModPhys.83.1467}}.
\newline\urlprefix\url{https://link.aps.org/doi/10.1103/RevModPhys.83.1467}

\bibitem{PhysRevLett.113.232502}
A.~Spyrou, S.~N. Liddick, A.~C. Larsen, M.~Guttormsen, K.~Cooper, A.~C. Dombos, D.~J. Morrissey, F.~Naqvi, G.~Perdikakis, S.~J. Quinn, T.~Renstr\o{}m, J.~A. Rodriguez, A.~Simon, C.~S. Sumithrarachchi, R.~G.~T. Zegers, \href{https://link.aps.org/doi/10.1103/PhysRevLett.113.232502}{Novel technique for constraining $r$-process ($n$, $\ensuremath{\gamma}$) reaction rates}, Phys. Rev. Lett. 113 (2014) 232502.
\newblock \href {https://doi.org/10.1103/PhysRevLett.113.232502} {\path{doi:10.1103/PhysRevLett.113.232502}}.
\newline\urlprefix\url{https://link.aps.org/doi/10.1103/PhysRevLett.113.232502}

\bibitem{PhysRevLett.116.242502}
S.~N. Liddick, A.~Spyrou, B.~P. Crider, F.~Naqvi, A.~C. Larsen, M.~Guttormsen, M.~Mumpower, R.~Surman, G.~Perdikakis, D.~L. Bleuel, A.~Couture, L.~Crespo~Campo, A.~C. Dombos, R.~Lewis, S.~Mosby, S.~Nikas, C.~J. Prokop, T.~Renstrom, B.~Rubio, S.~Siem, S.~J. Quinn, \href{https://link.aps.org/doi/10.1103/PhysRevLett.116.242502}{Experimental neutron capture rate constraint far from stability}, Phys. Rev. Lett. 116 (2016) 242502.
\newblock \href {https://doi.org/10.1103/PhysRevLett.116.242502} {\path{doi:10.1103/PhysRevLett.116.242502}}.
\newline\urlprefix\url{https://link.aps.org/doi/10.1103/PhysRevLett.116.242502}

\bibitem{PhysRevLett.117.142701}
A.~Spyrou, S.~N. Liddick, F.~Naqvi, B.~P. Crider, A.~C. Dombos, D.~L. Bleuel, B.~A. Brown, A.~Couture, L.~Crespo~Campo, M.~Guttormsen, A.~C. Larsen, R.~Lewis, P.~M\"oller, S.~Mosby, M.~R. Mumpower, G.~Perdikakis, C.~J. Prokop, T.~Renstr\o{}m, S.~Siem, S.~J. Quinn, S.~Valenta, \href{https://link.aps.org/doi/10.1103/PhysRevLett.117.142701}{Strong neutron-$\ensuremath{\gamma}$ competition above the neutron threshold in the decay of $^{70}\mathrm{Co}$}, Phys. Rev. Lett. 117 (2016) 142701.
\newblock \href {https://doi.org/10.1103/PhysRevLett.117.142701} {\path{doi:10.1103/PhysRevLett.117.142701}}.
\newline\urlprefix\url{https://link.aps.org/doi/10.1103/PhysRevLett.117.142701}

\bibitem{Misch2024}
G.~W. Misch, M.~R. Mumpower, \href{https://doi.org/10.1140/epjs/s11734-024-01136-z}{Astromers: status and prospects}, The European Physical Journal Special Topics 233~(5) (2024) 1075--1099.
\newblock \href {https://doi.org/10.1140/epjs/s11734-024-01136-z} {\path{doi:10.1140/epjs/s11734-024-01136-z}}.
\newline\urlprefix\url{https://doi.org/10.1140/epjs/s11734-024-01136-z}

\bibitem{GOLDANSKII1976393}
V.~Goldanskii, V.~Namiot, \href{https://www.sciencedirect.com/science/article/pii/0370269376906651}{On the excitation of isomeric nuclear levels by laser radiation through inverse internal electron conversion}, Physics Letters B 62~(4) (1976) 393--394.
\newblock \href {https://doi.org/https://doi.org/10.1016/0370-2693(76)90665-1} {\path{doi:https://doi.org/10.1016/0370-2693(76)90665-1}}.
\newline\urlprefix\url{https://www.sciencedirect.com/science/article/pii/0370269376906651}

\bibitem{Chiara2018}
C.~J. Chiara, J.~J. Carroll, M.~P. Carpenter, J.~P. Greene, D.~J. Hartley, R.~V.~F. Janssens, G.~J. Lane, J.~C. Marsh, D.~A. Matters, M.~Polasik, J.~Rzadkiewicz, D.~Seweryniak, S.~Zhu, S.~Bottoni, A.~B. Hayes, S.~A. Karamian, \href{https://doi.org/10.1038/nature25483}{Isomer depletion as experimental evidence of nuclear excitation by electron capture}, Nature 554~(7691) (2018) 216--218.
\newblock \href {https://doi.org/10.1038/nature25483} {\path{doi:10.1038/nature25483}}.
\newline\urlprefix\url{https://doi.org/10.1038/nature25483}

\bibitem{Carroll2024}
J.~J. Carroll, C.~J. Chiara, \href{https://doi.org/10.1140/epjs/s11734-024-01149-8}{Isomer depletion}, The European Physical Journal Special Topics 233~(5) (2024) 1151--1160.
\newblock \href {https://doi.org/10.1140/epjs/s11734-024-01149-8} {\path{doi:10.1140/epjs/s11734-024-01149-8}}.
\newline\urlprefix\url{https://doi.org/10.1140/epjs/s11734-024-01149-8}

\bibitem{BOLOTSKIKH2011146}
P.~Bolotskikh, D.~Grinfeld, A.~Makarov, M.~Monastyrskiy, \href{https://www.sciencedirect.com/science/article/pii/S0168900210029724}{Coulomb dynamics of ion bunches in multi-reflection electrostatic traps}, Nuclear Instruments and Methods in Physics Research Section A: Accelerators, Spectrometers, Detectors and Associated Equipment 645~(1) (2011) 146--152, the Eighth International Conference on Charged Particle Optics.
\newblock \href {https://doi.org/https://doi.org/10.1016/j.nima.2010.12.170} {\path{doi:https://doi.org/10.1016/j.nima.2010.12.170}}.
\newline\urlprefix\url{https://www.sciencedirect.com/science/article/pii/S0168900210029724}

\bibitem{RosenbuschAIP2013}
M.~Rosenbusch, S.~Kemnitz, R.~Schneider, L.~Schweikhard, R.~Tschiersch, R.~N. Wolf, \href{https://aip.scitation.org/doi/abs/10.1063/1.4796061}{Towards systematic investigations of space-charge phenomena in multi-reflection ion traps}, AIP Conference Proceedings 1521~(1) (2013) 53--62.
\newblock \href {http://arxiv.org/abs/https://aip.scitation.org/doi/pdf/10.1063/1.4796061} {\path{arXiv:https://aip.scitation.org/doi/pdf/10.1063/1.4796061}}, \href {https://doi.org/10.1063/1.4796061} {\path{doi:10.1063/1.4796061}}.
\newline\urlprefix\url{https://aip.scitation.org/doi/abs/10.1063/1.4796061}

\bibitem{RosenbuschAIP2015}
M.~Rosenbusch, P.~Chauveau, P.~Delahaye, G.~Marx, L.~Schweikhard, F.~Wienholtz, R.~N. Wolf, \href{https://aip.scitation.org/doi/abs/10.1063/1.4923120}{Delayed bunching for multi-reflection time-of-flight mass separation}, AIP Conference Proceedings 1668~(1) (2015) 050001.
\newblock \href {http://arxiv.org/abs/https://aip.scitation.org/doi/pdf/10.1063/1.4923120} {\path{arXiv:https://aip.scitation.org/doi/pdf/10.1063/1.4923120}}, \href {https://doi.org/10.1063/1.4923120} {\path{doi:10.1063/1.4923120}}.
\newline\urlprefix\url{https://aip.scitation.org/doi/abs/10.1063/1.4923120}

\bibitem{PhysRevE.104.065202}
D.~Gupta, R.~Singh, R.~Ringle, C.~R. Nicoloff, I.~Rahinov, O.~Heber, D.~Zajfman, \href{https://link.aps.org/doi/10.1103/PhysRevE.104.065202}{Particle-in-cell techniques for the study of space charge effects in an electrostatic ion beam trap}, Phys. Rev. E 104 (2021) 065202.
\newblock \href {https://doi.org/10.1103/PhysRevE.104.065202} {\path{doi:10.1103/PhysRevE.104.065202}}.
\newline\urlprefix\url{https://link.aps.org/doi/10.1103/PhysRevE.104.065202}

\bibitem{REITER2020431}
M.~Reiter, F.~Ames, C.~Andreoiu, S.~{Ayet San Andrés}, C.~Babcock, B.~Barquest, J.~Bergmann, J.~Bollig, T.~Brunner, T.~Dickel, J.~Dilling, I.~Dillmann, E.~Dunling, A.~Finlay, G.~Gwinner, L.~Graham, C.~Hornung, B.~Kootte, R.~Klawitter, P.~Kunz, Y.~Lan, D.~Lascar, J.~Lassen, E.~Leistenschneider, R.~Li, J.~McKay, M.~Mostamand, S.~Paul, W.~Plaß, C.~Scheidenberger, B.~Schultz, R.~Steinbrügge, A.~Teigelhoefer, R.~Thompson, M.~Wieser, C.~Will, A.~Kwiatkowski, \href{https://www.sciencedirect.com/science/article/pii/S0168583X19302186}{{Improved beam diagnostics and optimization at ISAC via TITAN’s MR-TOF-MS}}, Nuclear Instruments and Methods in Physics Research Section B: Beam Interactions with Materials and Atoms 463 (2020) 431--436.
\newblock \href {https://doi.org/https://doi.org/10.1016/j.nimb.2019.04.034} {\path{doi:https://doi.org/10.1016/j.nimb.2019.04.034}}.
\newline\urlprefix\url{https://www.sciencedirect.com/science/article/pii/S0168583X19302186}

\bibitem{AU2023375}
M.~Au, M.~Athanasakis-Kaklamanakis, L.~Nies, J.~Ballof, R.~Berger, K.~Chrysalidis, P.~Fischer, R.~Heinke, J.~Johnson, U.~Köster, D.~Leimbach, B.~Marsh, M.~Mougeot, B.~Reich, J.~Reilly, E.~Reis, M.~Schlaich, C.~Schweiger, L.~Schweikhard, S.~Stegemann, J.~Wessolek, F.~Wienholtz, S.~Wilkins, W.~Wojtaczka, C.~Düllmann, S.~Rothe, \href{https://www.sciencedirect.com/science/article/pii/S0168583X23002112}{In-source and in-trap formation of molecular ions in the actinide mass range at cern-isolde}, Nuclear Instruments and Methods in Physics Research Section B: Beam Interactions with Materials and Atoms 541 (2023) 375--379.
\newblock \href {https://doi.org/https://doi.org/10.1016/j.nimb.2023.05.015} {\path{doi:https://doi.org/10.1016/j.nimb.2023.05.015}}.
\newline\urlprefix\url{https://www.sciencedirect.com/science/article/pii/S0168583X23002112}

\bibitem{PhysRevC.107.064604}
M.~Au, M.~Athanasakis-Kaklamanakis, L.~Nies, R.~Heinke, K.~Chrysalidis, U.~K\"oster, P.~Kunz, B.~Marsh, M.~Mougeot, L.~Schweikhard, S.~Stegemann, Y.~Vila~Gracia, C.~E. D\"ullmann, S.~Rothe, \href{https://link.aps.org/doi/10.1103/PhysRevC.107.064604}{{Production of neptunium and plutonium nuclides from uranium carbide using 1.4-GeV protons}}, Phys. Rev. C 107 (2023) 064604.
\newblock \href {https://doi.org/10.1103/PhysRevC.107.064604} {\path{doi:10.1103/PhysRevC.107.064604}}.
\newline\urlprefix\url{https://link.aps.org/doi/10.1103/PhysRevC.107.064604}

\bibitem{HORNUNG2023257}
C.~Hornung, T.~Dickel, D.~Amanbayev, S.~{Ayet San Andrés}, D.~L. Balabanski, S.~Beck, J.~Bergmann, P.~Constantin, J.~Ebert, H.~Geissel, F.~Greiner, L.~Gröf, E.~Haettner, M.~N. Harakeh, J.-P. Hucka, N.~Kalantar-Nayestanaki, D.~A. Kostyleva, G.~Kripko-Koncz, I.~Miskun, A.~Mollaebrahimi, I.~Mukha, G.~Münzenberg, S.~Pietri, W.~R. Plaß, S.~Purushotaman, M.~P. Reiter, A.-K. Rink, H.~Roesch, C.~Scheidenberger, A.~Spătaru, Y.~K. Tanaka, H.~Weick, J.~Zhao, \href{https://www.sciencedirect.com/science/article/pii/S0168583X23001763}{Mass tagging: Verification and calibration of particle identification by high-resolution mass measurements}, Nuclear Instruments and Methods in Physics Research Section B: Beam Interactions with Materials and Atoms 541 (2023) 257--259.
\newblock \href {https://doi.org/https://doi.org/10.1016/j.nimb.2023.04.045} {\path{doi:https://doi.org/10.1016/j.nimb.2023.04.045}}.
\newline\urlprefix\url{https://www.sciencedirect.com/science/article/pii/S0168583X23001763}

\bibitem{BARQUEST201718}
B.~Barquest, G.~Bollen, P.~Mantica, K.~Minamisono, R.~Ringle, S.~Schwarz, C.~Sumithrarachchi, \href{http://www.sciencedirect.com/science/article/pii/S0168900217305892}{{RFQ beam cooler and buncher for collinear laser spectroscopy of rare isotopes}}, Nuclear Instruments and Methods in Physics Research Section A: Accelerators, Spectrometers, Detectors and Associated Equipment 866 (2017) 18 -- 28.
\newblock \href {https://doi.org/https://doi.org/10.1016/j.nima.2017.05.036} {\path{doi:https://doi.org/10.1016/j.nima.2017.05.036}}.
\newline\urlprefix\url{http://www.sciencedirect.com/science/article/pii/S0168900217305892}

\bibitem{Lapierre_2024}
A.~Lapierre, H.-J. Son, R.~Ringle, S.~Schwarz, A.~C.~C. Villari, \href{https://dx.doi.org/10.1088/1742-6596/2743/1/012063}{{High-Current Capability and Upgrades of the EBIS/T Charge-Breeding System in the Reaccelerator at the Facility for Rare-Isotope Beams}}, Journal of Physics: Conference Series 2743~(1) (2024) 012063.
\newblock \href {https://doi.org/10.1088/1742-6596/2743/1/012063} {\path{doi:10.1088/1742-6596/2743/1/012063}}.
\newline\urlprefix\url{https://dx.doi.org/10.1088/1742-6596/2743/1/012063}

\bibitem{doi:10.1007/s13361-017-1617-z}
T.~Dickel, W.~R. Plaß, W.~Lippert, J.~Lang, M.~I. Yavor, H.~Geissel, C.~Scheidenberger, \href{https://doi.org/10.1007/s13361-017-1617-z}{Isobar separation in a multiple-reflection time-of-flight mass spectrometer by mass-selective re-trapping}, Journal of the American Society for Mass Spectrometry 28~(6) (2017) 1079--1090, pMID: 28299713.
\newblock \href {http://arxiv.org/abs/https://doi.org/10.1007/s13361-017-1617-z} {\path{arXiv:https://doi.org/10.1007/s13361-017-1617-z}}, \href {https://doi.org/10.1007/s13361-017-1617-z} {\path{doi:10.1007/s13361-017-1617-z}}.
\newline\urlprefix\url{https://doi.org/10.1007/s13361-017-1617-z}

\bibitem{PascalsPaultrap}
T.~Fowler-Davis, et~al, In preparation (2025).

\bibitem{Kersevan:IPAC2019-TUPMP037}
R.~Kersevan, M.~Ady, \href{http://jacow.org/ipac2019/papers/tupmp037.pdf}{{R}ecent {D}evelopments of {M}onte{-C}arlo {C}odes {M}olflow+ and {S}ynrad+}, in: Proc. 10th International Particle Accelerator Conference (IPAC'19), Melbourne, Australia, 19-24 May 2019, no.~10 in International Particle Accelerator Conference, JACoW Publishing, Geneva, Switzerland, 2019, pp. 1327--1330, https://doi.org/10.18429/JACoW-IPAC2019-TUPMP037.
\newblock \href {https://doi.org/doi:10.18429/JACoW-IPAC2019-TUPMP037} {\path{doi:doi:10.18429/JACoW-IPAC2019-TUPMP037}}.
\newline\urlprefix\url{http://jacow.org/ipac2019/papers/tupmp037.pdf}

\bibitem{Sels2022}
S.~Sels, F.~M. Maier, M.~Au, P.~Fischer, C.~Kanitz, V.~Lagaki, S.~Lechner, E.~Leistenschneider, D.~Leimbach, E.~M. Lykiardopoulou, A.~A. Kwiatkowski, T.~Manovitz, Y.~N. Vila~Gracia, G.~Neyens, P.~Plattner, S.~Rothe, L.~Schweikhard, M.~Vilen, R.~N. Wolf, S.~Malbrunot-Ettenauer, \href{https://link.aps.org/doi/10.1103/PhysRevResearch.4.033229}{Doppler and sympathetic cooling for the investigation of short-lived radioactive ions}, Phys. Rev. Research 4 (2022) 033229.
\newblock \href {https://doi.org/10.1103/PhysRevResearch.4.033229} {\path{doi:10.1103/PhysRevResearch.4.033229}}.
\newline\urlprefix\url{https://link.aps.org/doi/10.1103/PhysRevResearch.4.033229}

\bibitem{YAVOR2009283}
M.~Yavor, \href{https://www.sciencedirect.com/science/article/pii/S1076567009016085}{Chapter 8 time-of-flight mass analyzers}, in: Optics of Charged Particle Analyzers, Vol. 157 of Advances in Imaging and Electron Physics, Elsevier, 2009, pp. 283--316.
\newblock \href {https://doi.org/https://doi.org/10.1016/S1076-5670(09)01608-5} {\path{doi:https://doi.org/10.1016/S1076-5670(09)01608-5}}.
\newline\urlprefix\url{https://www.sciencedirect.com/science/article/pii/S1076567009016085}

\bibitem{LECHNER2024169471}
S.~Lechner, S.~Sels, I.~Belosevic, F.~Buchinger, P.~Fischer, C.~Kanitz, V.~Lagaki, F.~Maier, P.~Plattner, L.~Schweikhard, M.~Vilen, S.~Malbrunot-Ettenauer, \href{https://www.sciencedirect.com/science/article/pii/S0168900224003978}{{Simulations of a cryogenic, buffer-gas filled Paul trap for low-emittance ion bunches}}, Nuclear Instruments and Methods in Physics Research Section A: Accelerators, Spectrometers, Detectors and Associated Equipment 1065 (2024) 169471.
\newblock \href {https://doi.org/https://doi.org/10.1016/j.nima.2024.169471} {\path{doi:https://doi.org/10.1016/j.nima.2024.169471}}.
\newline\urlprefix\url{https://www.sciencedirect.com/science/article/pii/S0168900224003978}

\bibitem{Klink_2024}
C.~Klink, M.~Schlaich, J.~Fischer, A.~Obertelli, A.~Schmidt, F.~Wienholtz, \href{https://dx.doi.org/10.1088/1748-0221/19/11/T11009}{{Development and commissioning of ion-optical elements for ion and antiproton beams with energies up to 5 keV}}, Journal of Instrumentation 19~(11) (2024) T11009.
\newblock \href {https://doi.org/10.1088/1748-0221/19/11/T11009} {\path{doi:10.1088/1748-0221/19/11/T11009}}.
\newline\urlprefix\url{https://dx.doi.org/10.1088/1748-0221/19/11/T11009}

\bibitem{MAIER2023167927}
F.~M. Maier, M.~Vilen, I.~Belosevic, F.~Buchinger, C.~Kanitz, S.~Lechner, E.~Leistenschneider, W.~Nörtershäuser, P.~Plattner, L.~Schweikhard, S.~Sels, F.~Wienholtz, S.~Malbrunot-Ettenauer, \href{https://www.sciencedirect.com/science/article/pii/S0168900222012190}{{Simulation studies of a 30-keV MR-ToF device for highly sensitive collinear laser spectroscopy}}, Nuclear Instruments and Methods in Physics Research Section A: Accelerators, Spectrometers, Detectors and Associated Equipment 1048 (2023) 167927.
\newblock \href {https://doi.org/https://doi.org/10.1016/j.nima.2022.167927} {\path{doi:https://doi.org/10.1016/j.nima.2022.167927}}.
\newline\urlprefix\url{https://www.sciencedirect.com/science/article/pii/S0168900222012190}

\bibitem{MAIER2025170365}
F.~Maier, F.~Buchinger, B.~A. Costa, H.~Heylen, C.~Kanitz, A.~Kwiatkowski, V.~Lagaki, S.~Lechner, E.~Leistenschneider, G.~Neyens, W.~Nörtershäuser, P.~Plattner, M.~Rosenbusch, L.~Schweikhard, S.~Malbrunot-Ettenauer, \href{https://www.sciencedirect.com/science/article/pii/S0168900225001664}{{Impact of the drift length on the performance of MR-ToF devices}}, Nuclear Instruments and Methods in Physics Research Section A: Accelerators, Spectrometers, Detectors and Associated Equipment 1075 (2025) 170365.
\newblock \href {https://doi.org/https://doi.org/10.1016/j.nima.2025.170365} {\path{doi:https://doi.org/10.1016/j.nima.2025.170365}}.
\newline\urlprefix\url{https://www.sciencedirect.com/science/article/pii/S0168900225001664}

\bibitem{WOLF20128}
R.~N. Wolf, G.~Marx, M.~Rosenbusch, L.~Schweikhard, \href{https://www.sciencedirect.com/science/article/pii/S1387380611004775}{Static-mirror ion capture and time focusing for electrostatic ion-beam traps and multi-reflection time-of-flight mass analyzers by use of an in-trap potential lift}, International Journal of Mass Spectrometry 313 (2012) 8--14.
\newblock \href {https://doi.org/https://doi.org/10.1016/j.ijms.2011.12.006} {\path{doi:https://doi.org/10.1016/j.ijms.2011.12.006}}.
\newline\urlprefix\url{https://www.sciencedirect.com/science/article/pii/S1387380611004775}

\bibitem{Fischer20182}
P.~Fischer, S.~Knauer, G.~Marx, L.~Schweikhard, \href{https://doi.org/10.1063/1.5009167}{In-depth study of in-trap high-resolution mass separation by transversal ion ejection from a multi-reflection time-of-flight device}, Review of Scientific Instruments 89~(1) (2018) 015114.
\newblock \href {http://arxiv.org/abs/https://doi.org/10.1063/1.5009167} {\path{arXiv:https://doi.org/10.1063/1.5009167}}, \href {https://doi.org/10.1063/1.5009167} {\path{doi:10.1063/1.5009167}}.
\newline\urlprefix\url{https://doi.org/10.1063/1.5009167}

\bibitem{doi:10.1021/acs.analchem.7b02797}
E.~T. Dziekonski, J.~T. Johnson, K.~W. Lee, S.~A. McLuckey, \href{https://doi.org/10.1021/acs.analchem.7b02797}{Fourier-transform ms and closed-path multireflection time-of-flight ms using an electrostatic linear ion trap}, Analytical Chemistry 89~(20) (2017) 10965--10972, pMID: 28926221.
\newblock \href {http://arxiv.org/abs/https://doi.org/10.1021/acs.analchem.7b02797} {\path{arXiv:https://doi.org/10.1021/acs.analchem.7b02797}}, \href {https://doi.org/10.1021/acs.analchem.7b02797} {\path{doi:10.1021/acs.analchem.7b02797}}.
\newline\urlprefix\url{https://doi.org/10.1021/acs.analchem.7b02797}

\bibitem{SimIon}
D.~Manura, D.~Dahl, \href{https://simion.com/}{Simion 8.1 user manual} (2008).
\newline\urlprefix\url{https://simion.com/}

\bibitem{RINGLE201142}
R.~Ringle, \href{https://www.sciencedirect.com/science/article/pii/S1387380610004872}{{3DCylPIC—A 3D particle-in-cell code in cylindrical coordinates for space charge simulations of ion trap and ion transport devices}}, International Journal of Mass Spectrometry 303~(1) (2011) 42--50.
\newblock \href {https://doi.org/https://doi.org/10.1016/j.ijms.2010.12.015} {\path{doi:https://doi.org/10.1016/j.ijms.2010.12.015}}.
\newline\urlprefix\url{https://www.sciencedirect.com/science/article/pii/S1387380610004872}

\bibitem{DICKEL20171}
T.~Dickel, M.~I. Yavor, J.~Lang, W.~R. Pla{\ss}, W.~Lippert, H.~Geissel, C.~Scheidenberger, \href{https://www.sciencedirect.com/science/article/pii/S1387380616302664}{Dynamical time focus shift in multiple-reflection time-of-flight mass spectrometers}, International Journal of Mass Spectrometry 412 (2017) 1--7.
\newblock \href {https://doi.org/https://doi.org/10.1016/j.ijms.2016.11.005} {\path{doi:https://doi.org/10.1016/j.ijms.2016.11.005}}.
\newline\urlprefix\url{https://www.sciencedirect.com/science/article/pii/S1387380616302664}

\bibitem{10.1063/5.0218649}
K.~König, F.~Köhler, J.~Palmes, H.~Badura, A.~Dockery, K.~Minamisono, J.~Meisner, P.~Müller, W.~Nörtershäuser, S.~Passon, \href{https://doi.org/10.1063/5.0218649}{{High voltage determination and stabilization for collinear laser spectroscopy applications}}, Review of Scientific Instruments 95~(8) (2024) 083307.
\newblock \href {http://arxiv.org/abs/https://pubs.aip.org/aip/rsi/article-pdf/doi/10.1063/5.0218649/20126153/083307\_1\_5.0218649.pdf} {\path{arXiv:https://pubs.aip.org/aip/rsi/article-pdf/doi/10.1063/5.0218649/20126153/083307\_1\_5.0218649.pdf}}, \href {https://doi.org/10.1063/5.0218649} {\path{doi:10.1063/5.0218649}}.
\newline\urlprefix\url{https://doi.org/10.1063/5.0218649}

\bibitem{PASSON2025101818}
S.~Passon, K.~König, F.~Schilling, B.~Maaß, J.~Meisner, W.~Nörtershäuser, \href{https://www.sciencedirect.com/science/article/pii/S2665917425000121}{{Ultra-stable 3D-Printed precision voltage divider for calibrations and experiments}}, Measurement: Sensors (2025) 101818\href {https://doi.org/https://doi.org/10.1016/j.measen.2025.101818} {\path{doi:https://doi.org/10.1016/j.measen.2025.101818}}.
\newline\urlprefix\url{https://www.sciencedirect.com/science/article/pii/S2665917425000121}

\bibitem{Wienholtzcom}
{Personal Communications with Frank Wienholtz, ISOLTRAP collaboration,} (2018).

\bibitem{BouwmanSummerStudent}
N.~Bouwman, {Identification and space charge studies of offline ion sources in ISOLTRAP.}, Summer Student Report, CERN (2023).

\bibitem{PhysRevC.103.025811}
C.~Izzo, J.~Bergmann, K.~A. Dietrich, E.~Dunling, D.~Fusco, A.~Jacobs, B.~Kootte, G.~Kripk\'o-Koncz, Y.~Lan, E.~Leistenschneider, E.~M. Lykiardopoulou, I.~Mukul, S.~F. Paul, M.~P. Reiter, J.~L. Tracy, C.~Andreoiu, T.~Brunner, T.~Dickel, J.~Dilling, I.~Dillmann, G.~Gwinner, D.~Lascar, K.~G. Leach, W.~R. Pla\ss{}, C.~Scheidenberger, M.~E. Wieser, A.~A. Kwiatkowski, \href{https://link.aps.org/doi/10.1103/PhysRevC.103.025811}{Mass measurements of neutron-rich indium isotopes for $r$-process studies}, Phys. Rev. C 103 (2021) 025811.
\newblock \href {https://doi.org/10.1103/PhysRevC.103.025811} {\path{doi:10.1103/PhysRevC.103.025811}}.
\newline\urlprefix\url{https://link.aps.org/doi/10.1103/PhysRevC.103.025811}

\bibitem{PhysRevC.104.065803}
S.~F. Paul, J.~Bergmann, J.~D. Cardona, K.~A. Dietrich, E.~Dunling, Z.~Hockenbery, C.~Hornung, C.~Izzo, A.~Jacobs, A.~Javaji, B.~Kootte, Y.~Lan, E.~Leistenschneider, E.~M. Lykiardopoulou, I.~Mukul, T.~Murb\"ock, W.~S. Porter, R.~Silwal, M.~B. Smith, J.~Ringuette, T.~Brunner, T.~Dickel, I.~Dillmann, G.~Gwinner, M.~MacCormick, M.~P. Reiter, H.~Schatz, N.~A. Smirnova, J.~Dilling, A.~A. Kwiatkowski, \href{https://link.aps.org/doi/10.1103/PhysRevC.104.065803}{{Mass measurements of $^{60-63}\mathrm{Ga}$ reduce x-ray burst model uncertainties and extend the evaluated $T=1$ isobaric multiplet mass equation}}, Phys. Rev. C 104 (2021) 065803.
\newblock \href {https://doi.org/10.1103/PhysRevC.104.065803} {\path{doi:10.1103/PhysRevC.104.065803}}.
\newline\urlprefix\url{https://link.aps.org/doi/10.1103/PhysRevC.104.065803}

\bibitem{YAVOR20151}
M.~I. Yavor, W.~R. Plaß, T.~Dickel, H.~Geissel, C.~Scheidenberger, \href{https://www.sciencedirect.com/science/article/pii/S1387380615000202}{Ion-optical design of a high-performance multiple-reflection time-of-flight mass spectrometer and isobar separator}, International Journal of Mass Spectrometry 381-382 (2015) 1--9.
\newblock \href {https://doi.org/https://doi.org/10.1016/j.ijms.2015.01.002} {\path{doi:https://doi.org/10.1016/j.ijms.2015.01.002}}.
\newline\urlprefix\url{https://www.sciencedirect.com/science/article/pii/S1387380615000202}

\bibitem{WIENHOLTZ2017285}
F.~Wienholtz, S.~Kreim, M.~Rosenbusch, L.~Schweikhard, R.~Wolf, \href{https://www.sciencedirect.com/science/article/pii/S1387380617301987}{Mass-selective ion ejection from multi-reflection time-of-flight devices via a pulsed in-trap lift}, International Journal of Mass Spectrometry 421 (2017) 285--293.
\newblock \href {https://doi.org/https://doi.org/10.1016/j.ijms.2017.07.016} {\path{doi:https://doi.org/10.1016/j.ijms.2017.07.016}}.
\newline\urlprefix\url{https://www.sciencedirect.com/science/article/pii/S1387380617301987}

\bibitem{Yoon2007}
O.~K. Yoon, I.~A. Zuleta, M.~D. Robbins, G.~K. Barbula, R.~N. Zare, \href{https://doi.org/10.1016/j.jasms.2007.07.030}{Simple template-based method to produce bradbury-nielsen gates}, Journal of the American Society for Mass Spectrometry 18~(11) (2007) 1901--1908.
\newblock \href {https://doi.org/10.1016/j.jasms.2007.07.030} {\path{doi:10.1016/j.jasms.2007.07.030}}.
\newline\urlprefix\url{https://doi.org/10.1016/j.jasms.2007.07.030}

\end{thebibliography}

\end{document}